%% file: main.tex
\documentclass{article} 
\usepackage{iclr2027/iclr2027_conference,times}

\input{iclr2027/math_commands.tex}

\usepackage[utf8]{inputenc} 
\usepackage[T1]{fontenc}    
\usepackage{hyperref}       
\usepackage{url}            
\usepackage{booktabs}       
\usepackage{amsfonts}       
\usepackage{nicefrac}       
\usepackage{microtype}      
\usepackage{xcolor}         

\usepackage{amssymb}
\usepackage{amsmath}
\usepackage{amsthm}
\newtheorem{theorem}{Theorem}

\usepackage{cancel}
\usepackage[ruled,vlined]{algorithm2e}
\usepackage{enumitem}
\usepackage{bm}
\usepackage{xcolor}
\usepackage{subfig}
\usepackage{graphicx}
\usepackage{listings}
\usepackage{multirow}
\usepackage{afterpage} 
\usepackage{makecell}
\usepackage{threeparttable}
\usepackage{adjustbox}
\usepackage{mathtools}
\usepackage{pifont}

\newcommand{\cmark}{\ding{51}}  
\newcommand{\xmark}{\ding{55}}  

\newcommand{\ie}{\textit{i.e.}}
\newcommand{\eg}{\textit{e.g.}}
\newcommand{\adj}{\tilde{A}}

\newcommand{\users}{\mathcal{U}}
\newcommand{\items}{\mathcal{I}}
\newcommand{\nodes}{\mathcal{N}}
\newcommand{\asplone}[1]{\underline{#1}}
\newcommand{\aspltwo}[1]{\textbf{#1}}
\newcommand{\asplthree}[1]{\underline{\textbf{#1}}}

\title{ASPIRE: Make Spectral Graph Collaborative Filtering Great Again via Adaptive Filter Learning}

\author{%
  Yunhang He$^{1}$ \qquad
  Cong Xu$^{1}$ \qquad
  Zhangchi Zhu$^{1}$ \qquad
  Hongzhi Yin$^{2}$ \qquad
  Wei Zhang$^{1,3}$\thanks{Corresponding author.}
  \\
  \\
  $^{1}$East China Normal University,
  $^{2}$The University of Queensland,
  $^{3}$Shanghai Innovation Institute \\
  \texttt{yhhe2004@gmail.com, congxueric@gmail.com, zczhu@stu.ecnu.edu.cn} \\
  \texttt{h.yin1@uq.edu.au, zhangwei.thu2011@gmail.com}
}

\iclrfinalcopy

\begin{document}

\maketitle

\begin{abstract}
  Graph filter design is central to spectral collaborative filtering, yet most existing methods rely on manually tuned hyperparameters rather than fully learnable filters. 
  We show that this challenge stems from a bias in traditional recommendation objectives, which induces a spectral phenomenon termed low-frequency explosion, thereby fundamentally hindering the effective learning of graph filters.
  To overcome this limitation, we propose a novel \underline{A}daptive \underline{S}pectral gra\underline{P}h collaborat\underline{I}ve filte\underline{R}ing fram\underline{E}work (ASPIRE) based on a bi-level optimization objective. 
  Guided by our theoretical analysis, we disentangle the filter learning objective, which in turn leads to excellent recommendation performance, spectral adaptivity, and training stability in practice.
  Extensive experiments show our learned filters match the performance of carefully engineered task-specific designs.
  Furthermore, ASPIRE is equally effective in LLM-powered collaborative filtering. 
  Our findings demonstrate that graph filter learning is viable and generalizable, paving the way for more expressive graph neural networks in collaborative filtering.
  Our code is available at https://github.com/yhhe2004/ASPIRE.
\end{abstract}

\section{Introduction}
\label{section-intro}

Graph Neural Networks (GNNs)~\citep{kipf2016gcn, bruna2013earlygnn} have become a significant paradigm in Collaborative Filtering (CF), where user-item interactions are modeled as bipartite graphs and representation propagation serves as implicit preference aggregation~\cite{gilmer2017neural, wang2019ngcf, ju2024tagcf}.
From a spectral perspective, many successful models can be interpreted as applying graph filters that emphasize specific frequency components of the interaction graph~\citep{shen2021gfcf, peng2022svdgcn, aiolli2013msd}.
This spectral view provides a principled understanding of why low-frequency smoothing improves recommendation accuracy and has inspired a series of filter-based designs~\citep{peng2024sgfcf, liu2023pgsp, zhu2024exploring}.

\afterpage{
    \clearpage
    \begin{figure}[]
      \centering
      \hspace*{0.0001\textwidth}
      \subfloat[(a) NDCG@20 Trajectory]{\includegraphics[width=0.325\textwidth]{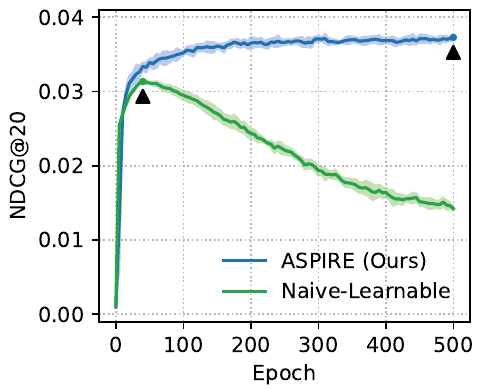}} \hfill
      \subfloat[(b) Naive-Learnable Filter Curve]{\includegraphics[width=0.325\textwidth]{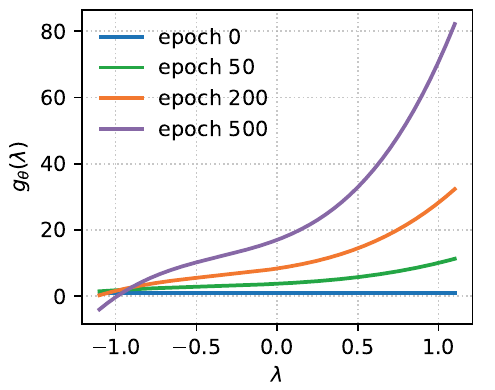}} \hfill
      \subfloat[(c) ASPIRE (Ours) Filter Curve]{\includegraphics[width=0.325\textwidth]{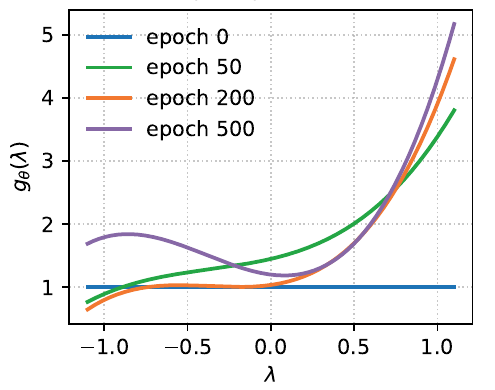}} 
      \hspace*{0.01\textwidth} 
    \caption{
        Comparison between ASPIRE and Naive-Learnable (a simple learnable polynomial filter) on the Baby dataset.
        In (a), $\blacktriangle$ marks the epoch achieving the best metric performance.
        In (b) and (c), we plot the learned filter $g_\theta(\lambda)$ in the spectral domain, whose effect is described by ~\eqref{equation-spectral-filtering}. 
        Note that the eigenvalues $\lambda_k$ of $\adj$ range from $-1$ (high-frequency) to $1$ (low-frequency). 
        Refer to Appendix~\ref{appendix-stability} for supplementary results.
    }
    \label{fig-intro-group}
    \end{figure}
}

Despite this theoretical clarity, a striking gap remains between expectation and practice.
Intuitively, learnable spectral filters should generalize well to diverse signal distributions with minimal hyperparameter tuning, mitigate over-smoothing~\citep{liu2021impgcn, wu2024afdgcf, li2018oversmoothing}, and enable more expressive graph neural networks.
However, empirical evidence consistently suggests otherwise.
Most existing spectral collaborative filtering methods rely on manually designed filters with carefully tuned hyperparameters, while learnable filters often exhibit unstable training behavior and inferior performance~\citep{he2025ssc} (see Figure~\ref{fig-intro-group}(a)).
This discrepancy raises a fundamental question:
\emph{Why do learnable spectral filters fail when trained with standard recommendation objectives?}
To better understand this phenomenon, we examine how the learned spectral response evolves during training.
Figure~\ref{fig-intro-group}(b) plots the filter curves at different training stages for a naive learnable spectral filter.
We observe a clear and consistent phenomenon termed \emph{Low-Frequency Explosion}: the magnitude assigned to low-frequency components grows continuously throughout training and never converges.

In this work, we show that this behavior is not accidental.
We first prove that the frequency bias of common recommendation objectives breaks \emph{Filter Stability} of learnable spectral modules.
It over-allocates optimization emphasis to low-frequency components, and systematically under-optimizes high-frequency signals, even when such signals are informative for recommendation~\citep{guo2023jgcf, choi2023bspm, liu2025simgcf}.
Consequently, directly learning spectral filters under conventional supervision leads to spectral imbalance and the collapse of \emph{Metric Stability}.

Building on the above theoretical analysis and insights from prior spectral analysis methods~\citep{guo2023jgcf, he2025ssc}, we propose ASPIRE, a bi-level optimization framework that decouples filter learning and directly optimizes for the fundamental goal of graph filters: predicting oracle interactions.
Our novel optimization objective completely alters the spectral learning dynamics.
As shown in Figure~\ref{fig-intro-group}(c), the filter learned under ASPIRE maintains a balanced spectral profile and is able to preserve informative high-frequency components.
In extensive experiments across multiple graph structures and evaluation settings, ASPIRE learns spectral filters whose performance matches carefully engineered, task-specific designs, while exhibiting markedly improved training stability.
Apart from validations on diverse graph scenarios, we further conduct experiments on LLM-powered CF task with two distinct adapter designs, consistently achieving outstanding performance and strong stability.

Our contributions are summarized as follows:
\begin{itemize}[leftmargin=*]
    \item We identify an intrinsic low-frequency bias in conventional recommendation objectives and provide a theoretical proof that aligns with the empirical observations in the spectral domain, thereby explaining why naive learnable filters collapse during training.
    \item We present a adaptive graph filter learning framework ASPIRE that separates filter optimization from the traditional objective, avoiding low-frequency explosion.
    \item We empirically demonstrate that our approach achieves comparable performance with handcrafted filters while significantly improving filter and metric stability.
\end{itemize}

\section{Preliminaries}

\subsection{Graph Collaborative Filtering}

Collaborative filtering is a fundamental paradigm in recommender systems, which aims to infer user preferences from historical user-item interaction data and generate personalized recommendation lists. 
In the implicit feedback setting, the interaction data is typically represented as a sparse binary matrix 
$R \in \mathbb{R}^{|\users| \times |\items|}$, 
where $r_{ui} \in \{0,1\}$ indicates whether user $u$ has interacted with item $i$.
The interaction data can also be viewed as a bipartite graph $\mathcal{G} = (\mathcal{E}, \mathcal{N})$ with $\nodes = \users \cup \items$, and its adjacency matrix $A$ can be derived from $R$.
In graph collaborative filtering, users and items are often associated with learnable embeddings $E \in \mathbb{R}^{|\nodes| \times d}$, where $d$ denotes the embedding dimension.
A common practice is to perform message propagation over the symmetrically normalized bipartite adjacency matrix $\tilde{A} = D^{-1/2} A D^{-1/2}$.
For personalized ranking optimization, Bayesian Personalized Ranking (BPR) loss~\citep{rendle2012bpr} is widely adopted in mainstream CF methods.

\subsection{Spectral Graph Collaborative Filtering}

Generally, spectral collaborative filtering~\citep{shen2021gfcf, qin2025polycf} methods use eigendecomposition to transform embeddings into the spectral domain, perform filtering on graph signals, and then transform them back to the original space.
Specifically, given a normalized bipartite adjacency matrix $\adj$, spectral filtering can be written as
\begin{equation}
\label{equation-spectral-filtering}
  H = U g(\Lambda) U^T E = U\,\mathrm{diag}\big(
      g(\lambda_1), \ldots, g(\lambda_{|\users| + |\items|})
  \big)\,U^T E,
\end{equation}
where \(U\) and \(\Lambda\) are eigenvectors and eigenvalues of \( \adj \) and \(g(\cdot)\) denotes a spectral filter applied to each frequency.
As a classic graph convolution-based CF method, LightGCN~\citep{he2020lightgcn} can also be interpreted from a spectral perspective as a simple uniform polynomial graph filter with $g(\lambda) = \frac{1}{L+1}\sum_{l=0}^{L}\lambda^{l}$:
\begin{equation}
  H = \frac{1}{L+1}\sum_{l=0}^{L} \tilde{A}^l E =  U \left ( \frac{1}{L+1} \sum_{l=0}^{L} \Lambda^l \right ) U^T E.
\label{equation-lightgcn-spectral-form}
\end{equation}
While some methods use the scoring matrix as the filter input~\citep{liu2023pgsp, qin2025polycf} and rely on matrix decomposition~\citep{peng2024sgfcf, kim2025chebycf}, our work targets graph embedding- and graph convolution-based filtering mechanisms.

\section{Analysis and Methodology}

\subsection{Understanding Low-Frequency Explosion from a Loss Perspective}

To explain the phenomenon observed in Figure~\ref{fig-intro-group}(b), we analyze the recommendation objective from a spectral perspective. 
Under the Perfect Uniformity condition defined by~\citet{wang2022directau}, and assuming that the mean absolute gradient deviation $\delta$ defined in~\eqref{equation-negligible-delta} is negligible, the following theorem characterizes how the BPR objective shapes individual spectral responses near convergence, providing insight into the observed training instability.

\begin{theorem}[]
\label{theorem-loss-bound}
Consider a spectral GNN near an optimal solution of the BPR loss, with the underlying graph $\phi(A) = \frac{1}{2|\mathcal{E}|}A = U \Lambda U^\top$.
Let $\rho \in (0,1)$ be the edge density of $A$, and define $\tilde{E} = U^\top E$, with $\tilde{E}_k$ denoting the $k$-th row of $\tilde{E}$.
For any eigenvalue $\lambda_k$ satisfying $|\lambda_k g(\lambda_k)| \, \|\tilde{E}_k\|^2 > 5(1 - \rho)\delta$,
the gradient of the BPR loss with respect to the spectral response $g(\lambda_k)$ satisfies
\[
\begin{dcases}
\frac{\partial \ell_{\mathrm{BPR}}}{\partial g(\lambda_k)}
< - \frac{1}{5(1-\rho)} \lambda_k g(\lambda_k)\|\tilde{E}_k\|^2 + \delta < 0, 
& \text{if } \lambda_k g(\lambda_k) > 0, \\[6pt]
\frac{\partial \ell_{\mathrm{BPR}}}{\partial g(\lambda_k)}
> - \frac{1}{5(1-\rho)} \lambda_k g(\lambda_k)\|\tilde{E}_k\|^2 - \delta > 0,
& \text{if } \lambda_k g(\lambda_k) < 0.
\end{dcases}
\]
\end{theorem}

The proof of Theorem~\ref{theorem-loss-bound}, along with empirical evidence supporting the Perfect Uniformity assumption, is provided in Appendix~\ref{appendix-proof-bpr}.
The proof proceeds by showing that positive sample pairs consistently promote low-frequency components, while negative pairs, under the Perfect Uniformity condition, exhibit a similar tendency.
Specifically, this theorem reveals that even when a model is close to an optimal solution, it still exerts a significant gradient on the filter $g_\theta(\Lambda)$, which undermines filter stability. 
Moreover, analyzing this bound helps identify the key factors driving the filter toward low-frequency explosion.

From ~\eqref{equation-posneg-filter} in the Appendix, when user and item embeddings are symmetrically filtered and predictions are computed via inner product, positive and negative $g(\lambda_k)$ are functionally equivalent, with only their absolute values impacting filtering.
Without loss of generality, we consider the common case where $g(\lambda_k) > 0$.
Under this condition, the optimization dynamics are directly modulated by the eigenvalue $\lambda_k$.
When $\lambda_k > 0$, the gradient drives $g(\lambda_k)$ to increase, whereas when $\lambda_k < 0$, the gradient pushes $g(\lambda_k)$ toward zero.
Moreover, the gradient magnitude contains a multiplicative factor $\lambda_k$.
When $g(\lambda_k)$ and $\|\tilde{E}_k\|^2$ are of comparable scale across different $k$, spectral components associated with larger $|\lambda_k|$ tend to receive stronger updates.
As optimization proceeds, this effect may further amplify $g(\lambda_k)$ in the low frequency area, leading to increasingly larger gradients and thus introducing a severe spectral bias during training.


Therefore, learnable graph filters suffer from an inherent strong low-pass bias and over-regularization, leading to severe high-frequency suppression and a prominent low-frequency explosion.
This lopsided frequency preference precisely aligns with the trend in Figure~\ref{fig-intro-group}(b).
Notably, the same phenomenon also occurs under the cross-entropy (CE) loss, with theoretical analysis provided in Appendix~\ref{appendix-proof-ce} and experimental details elaborated in Appendix~\ref{appendix-stab-ce}.

\subsection{Adaptive Graph Filter Learning}

As revealed in the previous section, conventional recommendation objectives are inherently misaligned with the purpose of graph filters. 
Building on insights from prior spectral analysis methods, we propose a bi-level optimization scheme to decouple filter learning.

\paragraph{Pre-filter Normalization.}

To prevent dominant high-norm embeddings from overwhelming the filter learning process,
we first perform row-wise $\ell_2$ normalization on the embedding matrix $\bar{E}_i=E_i / \|E_i\|_2$, where $E_i$ denotes the $i$-th row of $E$.
This operation removes scale variations across nodes before spectral filtering.
It accelerates optimization convergence, boosts recommendation performance, and greatly enhances training stability, as detailed in Appendix~\ref{appendix-effect-norm}.

\paragraph{Adaptive Filtering.}

Given an arbitrary graph $\mathcal{G}$ with its graph matrix factorized as $U \Lambda U^\top$, the adaptive filtering procedure on the input embeddings $\bar{E}$ can be written as
\begin{equation}
  H = \mathcal{F}\!\left(U g_\theta(\Lambda) U^\top \bar{E}\right), 
  \quad
  g_\theta(\Lambda) = \sum_{l=0}^{L} \theta_l \Lambda^{l},
  \label{equation-adaptive-filter}
\end{equation}
where $g_\theta(\Lambda)$ denotes a learnable spectral response function with learnable coefficients 
$\theta = \{\theta_l\}_{l=0}^{L}$, and $\mathcal{F}$ represents a post-filter operator that maps the filtered embeddings to the final representation space (\eg, multi-branch fusion).
We initialize the filter as a full-pass filter with $\theta_0 = 1$ and $\theta_{1\ldots L} = 0$, which corresponds to an identity transformation and provides a general starting point that allows the model to learn an appropriate graph filter from data.

\paragraph{Bi-level Optimization Formulation.}

Previous spectral analysis methods, either from a matrix transformation perspective~\citep{guo2023jgcf} or a signal importance perspective~\citep{he2025ssc}, share the same intuition that an effective filter should construct oracle interactions from prior knowledge.
Based on this insight, we reformulate the supervision signal and propose the following training objective:
\begin{equation}
\begin{gathered}
\theta^* = \argmin_{\theta} \mathcal{L}_{\mathrm{valid}}\left( \theta , E^*(\theta) \right), \\
\text{s.t.} \quad E^*(\theta) = \argmin_{E} \mathcal{L}_{\mathrm{train}}(\theta , E).
\end{gathered}
\end{equation}
This formulation is well-motivated, as it optimizes the filter parameters $\theta$ by minimizing the validation loss, which serves as an oracle with respect to the remaining model parameters.
In general CF tasks, these parameters correspond to the embedding $E$.

\paragraph{Bi-level Optimization Solution.}

\begin{table*}[t]
\centering
\caption{Comparison of recommendation performance and computational overhead for upper-level gradient approximations (memory in MB). `Avg-P' refers to the baseline filter defined in ~\eqref{equation-lightgcn-spectral-form}. `Ex' and `Im' denote the explicit and implicit term, respectively. Refer to Appendix~\ref{appendix-time-memory} for supplementary results.}
\label{table-cost-brief}
\scalebox{.85}{
\begin{tabular}{c | cccc | cccc}
\toprule
\multirow{2}{*}{Method} 
& \multicolumn{4}{c|}{Baby} 
& \multicolumn{4}{c}{Electronics} \\
& R@20 & N@20 & (Extra) Time & (Extra) Mem 
& R@20 & N@20 & (Extra) Time & (Extra) Mem \\
\midrule

Avg-P 
& 0.0851 & 0.0372 & 10m09s\phantom{00} & 488\phantom{00} 
& 0.0579 & 0.0272 & 120m16s\phantom{00} & 1970\phantom{00} \\

Ex\&Im 
& 0.0870 & 0.0380 & \quad+ 5m11s & \quad+ 148 
& 0.0636 & 0.0306 & \quad+ 161m14s & \quad+ 1082 \\

\textbf{Ex} 
& 0.0871 & 0.0380 & \quad+ 0m35s & \quad+ \phantom{0}\phantom{0}0 
& 0.0635 & 0.0306 & \quad+ \phantom{0}19m22s & \quad+ \phantom{0}\phantom{0}66 \\

\bottomrule
\end{tabular}
}
\end{table*}

Given that the filter $g_\theta(\Lambda)$ contains only a small number of parameters, we sample a subset of the validation data, which is considerably smaller than the training set, to construct an auxiliary set for upper-level optimization (see Appendix~\ref{appendix-sensitivity}).
Inspired by~\citet{yin2024dr4sr}, we employ a computationally efficient scheme that interleaves optimization across levels, executing a single upper-level update after $T$ steps of lower-level updates.
The lower-level optimization can be readily implemented via gradient descent with $\theta$ fixed.
The upper-level optimization involves two distinct sources of gradients: an explicit gradient from the filter parameter $\theta$, and an implicit gradient within $E^*$ inherited from the lower-level optimization.
\begin{equation}
\nabla_\theta \mathcal{L}_{\mathrm{valid}}\left( \theta , E^*(\theta) \right) = 
\underbrace{\nabla_\theta \mathcal{L}_{\mathrm{valid}}}_{\text{Explicit Term}} 
\underbrace{
    -
    \nabla_E \mathcal{L}_{\mathrm{valid}} \cdot 
    \left( \nabla_E^2 \mathcal{L}_{\mathrm{train}} \right)^{-1} \cdot 
    \nabla_\theta \nabla_E \mathcal{L}_{\mathrm{train}}
}_{\text{Implicit Term}}
\label{equation-neumann-approx}
\end{equation}
Our implicit differentiation follows the methods introduced by~\citet{lorraine2020implicitgrad, navon2020implicitgrad}. 
Inspired by prior first-order approximations in bi-level optimization and meta-learning~\citep{finn2017maml, liu2018darts, kwon2023f2sa}, we empirically examine the effect of the implicit term in Table~\ref{table-cost-brief}.
Results show that the implicit term contributes negligibly to this task, and we therefore omit it, leading to substantial reductions in both computational time and memory overhead.
We provide a thorough comparison of different approximation techniques in Appendix~\ref{appendix-time-memory}. 

\section{Empirical Evaluation}
\label{section-experiment}

To assess whether the learned graph filter exhibits desirable behaviors, we evaluate each method from three complementary perspectives.
We first briefly introduce experimental setups, with details in Appendix~\ref{appendix-exp-setup}.
All results in this paper are averaged over five runs.

\paragraph{Scenarios.}

To evaluate our method across different graph structures and model architectures, we consider three settings with different graphs $\mathcal{G}$ and post-filter operators $\mathcal{F}(\cdot)$: \emph{Homogeneous}, \emph{Heterogeneous}, and \emph{Dual-Branch}.
In the Homogeneous setting, we use the common interaction graph for $\mathcal{G}$ and set $\mathcal{F}(X)=X$.
In the Heterogeneous setting, we adopt the side information augmented graph $A_+$ proposed by~\citet{he2025ssc}, while keeping $\mathcal{F}(X)=X$.
In the Dual-Branch setting, inspired by~\citet{zhou2023freedom}, we employ both the interaction graph $\adj$ and the side information graph $S$, apply filtering to each branch separately, and fuse the resulting representations via average pooling, where $\mathcal{F}(\cdot)$ denotes this average pooling operation.

\paragraph{Datasets.}

We consider two types of side information: multimodal relationships and social connections.
Our preprocessing follows~\citet{zhou2023mmrec}, while graph construction follows~\citet{he2025ssc}.
For multimodal datasets, we use two widely used datasets, Amazon Baby and Electronics, with 5-core filtering.
The original modality data of each item, including both textual and visual modalities, are encoded into dense vectors using pretrained encoders. 
These representations are further transformed into a sparse item-item similarity graph via $k$-nearest-neighbors, where $k$ is empirically set to 10.
For social datasets, we also use two real-world datasets, Ciao and LastFM, with 3-core filtering.
The social connections are symmetrized to construct a user-user graph.
For both scenarios, we split the data into training, validation, and test sets with a ratio of 8:1:1.

\paragraph{Baselines.}

We consider three conventional filters that are commonly adopted as components in the aforementioned scenarios: Layer-$0$ (L-$0$), Layer-$n$ (L-$n$), and Average-Pooling (Avg-P).
Avg-P is especially popular, as it incorporates multi-order information from each layer~\citep{he2020lightgcn}.
We further implement two manually designed filters motivated by empirical observations in specific architectures: Jacobi-Basis (Jacobi) and Linear-Correction (Lin-C).
Jacobi is initially designed under Homogenerous setting~\citep{guo2023jgcf}, while Lin-C is designed for side-information augmented Heterogeneous graph~\citep{he2025ssc}.
Alongside the non-learnable but tunable filters above, we include Naive-Learnable (Naive-L) which simply adopts a learnable polynomial filter without extra designs. 
The corresponding spectral response function of each method is listed in Table~\ref{table-baseline}.
Appendix~\ref{appendix-baseline} presents representative models built with the above filters.

\subsection{Recommendation Performance}

We report the standard top-$N$ recommendation metrics Recall@$N$ and NDCG@$N$, where $N \in \{10, 20\}$.
\textbf{Since many baselines rely on carefully designed filters with extensively tuned hyperparameters, our goal is not necessarily to surpass these handcrafted filters in every case, but to achieve comparable performance while learning the filter adaptively.}

\begin{figure}[t]
\centering
\begin{minipage}[t]{0.35\textwidth}
\vspace{0pt}
\centering
\captionof{table}{Baseline filters functions.}
\label{table-baseline}

\scalebox{0.9}{
\renewcommand{\arraystretch}{1.6}
\begin{tabular}{c|c}
\toprule
  & $g(\lambda)$ \\
\midrule
L-$0$    & $1$ \\
L-$n$    & $\lambda^n$ \\
Avg-P    & $\frac{1}{L+1}\sum_{l=0}^{L}\lambda^{l}$ \\
Jacobi   & $\sum_{l=0}^{L}\beta_l \mathrm{Jacobi}_l(\lambda)$ \\
Lin-C    & $\frac{1}{L+1}\sum_{l=0}^{L}\mathrm{linear}(\lambda)^{l}$ \\
Naive-L  & $\sum_{l=0}^{L} \theta_l \lambda^{l}$ \\
\bottomrule
\end{tabular}
}
\end{minipage}
\hfill
\begin{minipage}[t]{0.64\textwidth}
\vspace{-2pt}
\centering
\includegraphics[width=\linewidth]{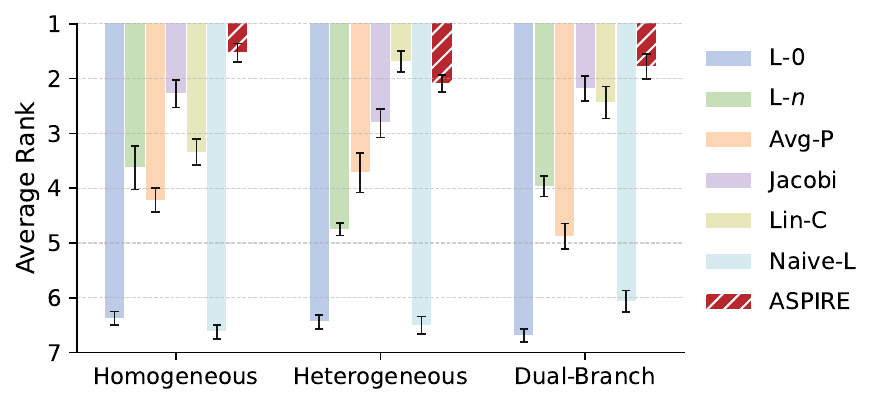}
\caption{Average rank (mean $\pm$ s.e.m.) of each filter across scenarios, computed from the results in Table~\ref{table-metric-comparison-main}.}
\label{figure-exp-avg-rank}
\end{minipage}
\end{figure}

\begin{table*}[t]
\centering
\caption{Performance comparison between ASPIRE and six baseline filter settings. $\Delta$ Avg-P denotes the percentage improvement over Avg-P. The leading zero before decimal points is omitted.}
\label{table-metric-comparison-main}
\scalebox{.67}{
\setlength{\tabcolsep}{1.12mm}
\begin{tabular}{c | c | cccc | cccc | cccc | cccc}
\toprule
& \multirow{2}{*}{Method} 
& \multicolumn{4}{c|}{Baby} 
& \multicolumn{4}{c|}{Electronics} 
& \multicolumn{4}{c|}{Ciao} 
& \multicolumn{4}{c}{LastFM} \\
& & R@10 & R@20 & N@10 & N@20 & R@10 & R@20 & N@10 & N@20 & R@10 & R@20 & N@10 & N@20 & R@10 & R@20 & N@10 & N@20 \\
\midrule

\multirow{8}{*}{\rotatebox{90}{Homogeneous}}
& L-$0$ & .0417 & .0638 & .0227 & .0284 & .0372 & .0557 & .0208 & .0256 & .0450 & .0737 & .0269 & .0353 & .2357 & .3220 & .2165 & .2513 \\
& L-$n$ & .0536 & .0831 & .0289 & .0365 & .0407 & .0600 & .0236 & .0286 & .0498 & .0784 & .0315 & .0398 & .2450 & .3316 & .2245 & .2595 \\
& Avg-P & .0543 & .0851 & .0293 & .0372 & .0393 & .0579 & .0224 & .0272 & .0473 & .0770 & .0295 & .0381 & .2527 & .3430 & .2326 & .2690 \\
& Jacobi & .0566 & .0873 & .0306 & .0385 & .0425 & .0618 & .0244 & .0294 & .0489 & .0783 & .0310 & .0395 & .2550 & .3425 & .2348 & .2700 \\
& Lin-C & .0542 & .0850 & .0292 & .0372 & .0394 & .0585 & .0225 & .0274 & .0484 & .0786 & .0303 & .0390 & .2558 & .3454 & .2352 & .2712 \\
& Naive-L & .0450 & .0690 & .0246 & .0307 & .0312 & .0463 & .0174 & .0213 & .0445 & .0678 & .0283 & .0352 & .2337 & .3241 & .2131 & .2496 \\
\\[-10pt]
\cline{2-18}
\\[-8pt]
& ASPIRE & \aspltwo{.0550} & \aspltwo{.0871} & \aspltwo{.0297} & \aspltwo{.0380} & \asplthree{.0440} & \asplthree{.0635} & \asplthree{.0255} & \asplthree{.0306} & .0489 & \asplthree{.0796} & .0309 & .0397 & \asplthree{.2565} & \asplthree{.3475} & \asplthree{.2366} & \asplthree{.2733} \\
& $\Delta$ Avg-P & 1.30\% & 2.38\% & 1.53\% & 2.19\% & 11.9\% & 9.72\% & 13.9\% & 12.4\% & 3.34\% & 3.44\% & 4.65\% & 4.25\% & 1.49\% & 1.30\% & 1.73\% & 1.58\% \\

\midrule\midrule

\multirow{8}{*}{\rotatebox{90}{Heterogeneous}}
& L-$0$ & .0417 & .0638 & .0227 & .0284 & .0372 & .0557 & .0208 & .0256 & .0450 & .0737 & .0269 & .0353 & .2357 & .3220 & .2165 & .2513 \\
& L-$n$ & .0576 & .0922 & .0312 & .0401 & .0432 & .0638 & .0242 & .0295 & .0517 & .0802 & .0326 & .0409 & .2435 & .3315 & .2238 & .2592 \\
& Avg-P & .0627 & .0990 & .0337 & .0430 & .0456 & .0675 & .0257 & .0313 & .0524 & .0808 & .0327 & .0410 & .2362 & .3257 & .2174 & .2536 \\
& Jacobi & .0654 & .0996 & .0358 & .0446 & .0436 & .0639 & .0248 & .0300 & .0554 & .0849 & .0347 & .0433 & .2541 & .3440 & .2346 & .2710 \\
& Lin-C & .0666 & .1044 & .0361 & .0458 & .0452 & .0677 & .0254 & .0312 & .0547 & .0861 & .0340 & .0431 & .2588 & .3502 & .2406 & .2772 \\
& Naive-L & .0538 & .0833 & .0293 & .0369 & .0330 & .0486 & .0187 & .0228 & .0442 & .0708 & .0282 & .0360 & .2332 & .3268 & .2128 & .2506 \\
\\[-10pt]
\cline{2-18}
\\[-8pt]
& ASPIRE & \aspltwo{.0666} & \aspltwo{.1027} & \aspltwo{.0361} & \aspltwo{.0454} & .0454 & \asplthree{.0679} & .0255 & .0313 & \asplone{.0540} & \asplone{.0844} & \asplone{.0338} & \asplone{.0426} & \aspltwo{.2570} & \aspltwo{.3476} & \aspltwo{.2389} & \aspltwo{.2755} \\
& $\Delta$ Avg-P & 6.22\% & 3.82\% & 7.17\% & 5.55\% & -0.50\% & 0.60\% & -0.79\% & -0.13\% & 3.08\% & 4.57\% & 3.22\% & 3.99\% & 8.79\% & 6.75\% & 9.88\% & 8.61\% \\

\midrule\midrule

\multirow{8}{*}{\rotatebox{90}{Dual-Branch}}
& L-$0$ & .0417 & .0638 & .0227 & .0284 & .0372 & .0557 & .0208 & .0256 & .0450 & .0737 & .0269 & .0353 & .2357 & .3220 & .2165 & .2513 \\
& L-$n$ & .0640 & .0985 & .0345 & .0433 & .0447 & .0665 & .0251 & .0307 & .0529 & .0827 & .0328 & .0414 & .2423 & .3286 & .2226 & .2573 \\
& Avg-P & .0630 & .0967 & .0343 & .0429 & .0465 & .0695 & .0260 & .0319 & .0519 & .0807 & .0326 & .0410 & .2360 & .3247 & .2171 & .2529 \\
& Jacobi & .0670 & .1029 & .0358 & .0450 & .0473 & .0698 & .0267 & .0325 & .0533 & .0826 & .0330 & .0416 & .2566 & .3471 & .2361 & .2726 \\
& Lin-C & .0642 & .0974 & .0347 & .0432 & .0465 & .0695 & .0260 & .0319 & .0547 & .0830 & .0339 & .0421 & .2578 & .3490 & .2393 & .2761 \\
& Naive-L & .0557 & .0866 & .0305 & .0384 & .0343 & .0508 & .0189 & .0231 & .0453 & .0702 & .0292 & .0365 & .2369 & .3258 & .2177 & .2536 \\
\\[-10pt]
\cline{2-18}
\\[-8pt]
& ASPIRE & \aspltwo{.0659} & \aspltwo{.1021} & \aspltwo{.0354} & \aspltwo{.0447} & \asplthree{.0479} & \asplthree{.0710} & \asplthree{.0270} & \asplthree{.0330} & .0525 & \aspltwo{.0830} & .0328 & \aspltwo{.0417} & \asplthree{.2580} & \asplthree{.3502} & \aspltwo{.2390} & \aspltwo{.2761} \\
& $\Delta$ Avg-P & 4.64\% & 5.61\% & 3.29\% & 4.08\% & 2.92\% & 2.12\% & 3.89\% & 3.29\% & 1.07\% & 2.77\% & 0.47\% & 1.72\% & 9.30\% & 7.84\% & 10.1\% & 9.16\% \\

\bottomrule
\end{tabular}
}

\end{table*}

\paragraph{Notations.}
Table~\ref{table-metric-comparison-main} demonstrates the overall performance comparison.
For a more detailed evaluation, we categorize the results into four levels.
Level~0 indicates results that outperform the initialization filter (\ie, L-$0$), which is satisfied by all reported results.
Level~1 indicates results that outperform all commonly used filters, marked with an \asplone{underline}.
Level~2 indicates results whose performance lies between the two manually designed filters, marked in \aspltwo{bold}.
Level~3 indicates results that outperform all manually designed filters, marked with both \asplthree{bold and underline}.
Note that some results may satisfy a higher level without satisfying Level~1; in such cases, we mark them as Level~0.

\paragraph{Observations.}
Across different scenarios and datasets, the manually designed filters - Lin-C and Jacobi - outperform empirical ones in most cases, likely due to their spectral insights and careful hyperparameter tuning. 
For ASPIRE, 75\% of the results reach at least Level 2, \ie, comparable to the two manually designed filters
Furthermore, 33.3\% of the results surpass both of these two manually designed filters. 
To intuitively compare the overall performance of all methods across scenarios, we plot their average rankings in Figure~\ref{figure-exp-avg-rank}. 
As shown in this figure, Jacobi and Lin-C achieve distinct advantages in their respective tailored scenarios (\ie, Homogeneous and Heterogeneous, respectively).
In contrast, ASPIRE delivers competitive and relatively stable performance across all settings, even outperforming both methods overall in the Homogeneous and Dual-Branch scenarios.
This clearly validates that our proposed learnable filter possesses stronger generalization ability.
By comparison, Naive-L performs significantly worse than ASPIRE due to the low-frequency explosion.
Compared with Avg-P, which is commonly used as a default component in many works, ASPIRE yields positive gains on the vast majority of metrics in various scenarios, with improvements up to 13.9\%.
This result shows that introducing ASPIRE as the graph filtering mechanism can generally enhance recommendation performance.
Notably, unlike manually designed filters, ASPIRE obviates the need for prior spectral analysis of the graph while achieving superior generalizability.

\subsection{Spectral Adaptivity}

\begin{figure}
  \includegraphics[width=0.99\textwidth]{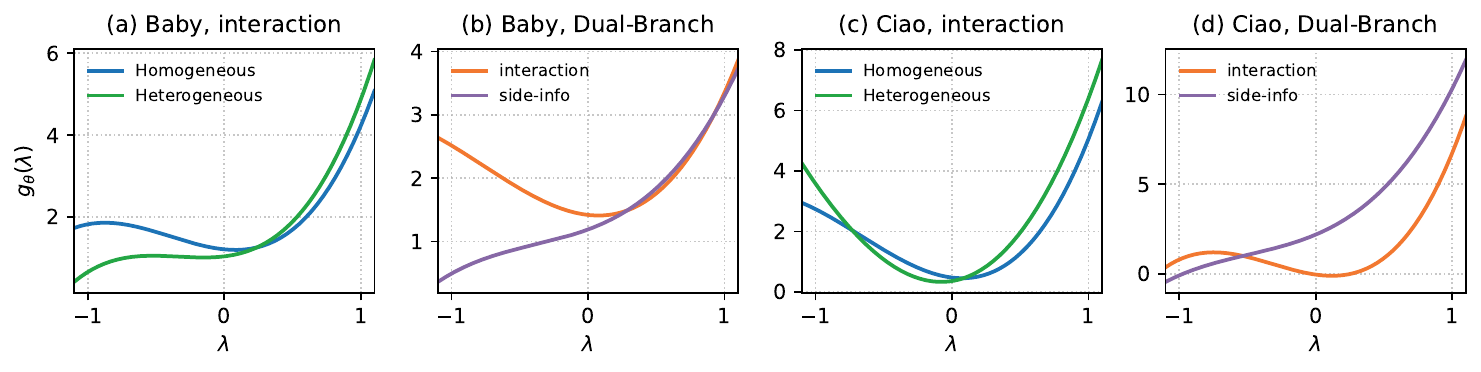}
  \caption{Learned filters across different graph settings. (a) and (c) present the filters for the interaction graph under the Homogeneous (standard) and Heterogeneous (augmented) settings, while (b) and (d) present the filters for the standard interaction graph and the side information graph in the Dual-Branch setting.}
  \label{figure-exp-filter-cmp}
\end{figure}

\textbf{A desirable filter should adapt to spectral variations across different graphs~\citep{luo2024scomgnn, rabiah2025gsprec}.}
We evaluate this property by examining whether the learned filters differ across datasets with distinct graph structures, and whether these differences are consistent with observations reported in prior studies.
The learned filters are visualized in Figure~\ref{figure-exp-filter-cmp}.

\textbullet~First, the learned filters vary noticeably across different graph types.
For example, unlike other graph types, the learned filter on the side-information graph (purple line) exhibits a clear low-pass tendency, which may explain why traditional low-pass filters perform well in similar scenarios~\citep{wang2023dsl, yu2025pgl}.
\enspace\textbullet~Second, prior studies suggest that effective recommendation models should properly emphasize both low- and high-frequency components of the interaction graph spectrum~\citep{kim2025chebycf, liu2025simgcf}.
In Figure~\ref{figure-exp-filter-cmp}, the blue and orange curves both exhibiting responses on low- and high-frequency regions.
Nevertheless, subtle differences are still observed, further supporting our first observation.
\enspace\textbullet~Additionally, by inspecting the oracle spectrum reported by~\citet{he2025ssc}, we observe that when the interaction graph is augmented with side information, the relative importance of low-frequency components increases compared to high-frequency ones.
Comparing the low-frequency regions in (a) and (c), the learned filters exhibit a similar trend (noting that the effective high-frequency spectrum shifts rightward away from $\lambda = -1$).

\subsection{Training Stability}
\label{section-stable-converge}

\begin{figure}[t]
\centering
\begin{minipage}[t]{0.72\textwidth}
\vspace{0pt}
\centering
\captionof{table}{Initialization Robustness. 
`E@x\%' indicates the initial epoch achieving x\% of the final N@20 on Baby. 
Evaluation is conducted every 5 epochs. 
Refer to Appendix~\ref{appendix-init-sensitivity} for supplementary results.}
\label{table-initialization-comparison}

\scalebox{0.74}{
\begin{tabular}{lccccccc}
\toprule
\multirow{2}{*}{Initialization} & \multirow{2}{*}{$g_\theta^{init}(\lambda)$} 
& \multicolumn{2}{c}{Convergence Speed} 
& \multicolumn{2}{c}{Final Perf - Baby}
& \multicolumn{2}{c}{Final Perf - Elec} \\
\cmidrule(lr){3-4} \cmidrule(lr){5-6} \cmidrule(lr){7-8}
& & E@80\% & E@95\% & R@20 & N@20 & R@20 & N@20 \\
\midrule
full-pass      & $1$ & 20  & 105 & 0.0871 & 0.0380 & 0.0635 & 0.0306 \\
zero-crossing  & $\lambda$ & 5   & 70  & 0.0865 & 0.0378 & 0.0635 & 0.0306 \\
low-pass       & $\frac{1}{4}\sum_{l=0}^{3}\lambda^{l}$ & 5 & 40 & 0.0884 & 0.0383 & 0.0634 & 0.0305 \\
mid-pass       & $1-\lambda^2$ & 100 & 145 & 0.0873 & 0.0381 & 0.0636 & 0.0307 \\
high-pass      & $1-2\lambda+\lambda^2$ & 105 & 175 & 0.0870 & 0.0375 & 0.0630 & 0.0304 \\
\midrule
\multicolumn{2}{c}{Baseline: Average-Pooling} & 60 & 190 & 0.0851 & 0.0372 & 0.0579 & 0.0272 \\
\bottomrule
\end{tabular}
}
\end{minipage}
\hfill
\begin{minipage}[t]{0.265\textwidth}
\vspace{0pt}
\centering
\includegraphics[width=\linewidth]{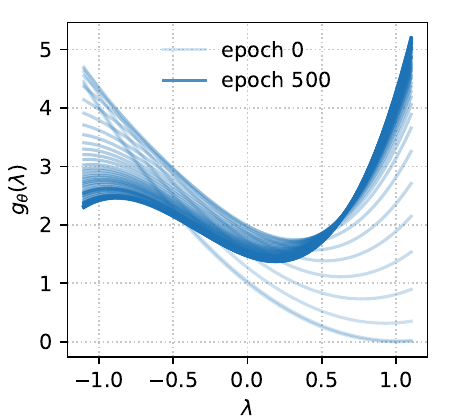}
\caption{Filter evolution of high-pass initialization. }
\label{figure-high-pass-evolution}
\end{minipage}
\end{figure}

Finally, we examine the Metric Stability and Filter Stability during training.
\textbf{A reliable method should exhibit stable metric trajectories, smooth parameter evolution, and consistent convergence under different initializations.}

As illustrated in Figure~\ref{fig-intro-group}, the performance of ASPIRE maintains a steady upward trend throughout training, in contrast to the sharp performance drop observed in the naive learnable model.
Meanwhile, the learned filter evolves smoothly and maintains a reasonable scale during the optimization process.
More frequent snapshots of the spectral response function are provided in Appendix~\ref{appendix-stability}.

We further investigate the performance under different filter initializations in Table~\ref{table-initialization-comparison}.
With the full-pass, zero-crossing, and low-pass initialization, the model converges faster and consistently outperforms the baseline Avg-P.
For mid-pass and high-pass initializations, which initially suppress low-frequency signals, the model experiences a longer warm-up stage but subsequently catches up and converges faster than Avg-P.
Although NDCG@20 of the high-pass initialization is not significantly better than Avg-P on Baby, its filter evolution (Figure~\ref{figure-high-pass-evolution}) shows that the learned filter eventually converges to the same solution obtained from the full-pass initialization.
In fact, across all the initialization settings above, the learned filters converge to essentially the same spectral shape (or its sign-flipped counterpart mirrored across the $x$-axis, which is functionally equivalent~\citep{liu2025simgcf}), with only slight differences in the resulting metrics.

\section{Extending ASPIRE to LLM-Powered Collaborative Filtering}
\label{section-llm}

\begin{figure}
  \centering
      \subfloat[(a) Illustrative Overview]{\includegraphics[width=0.7\textwidth]{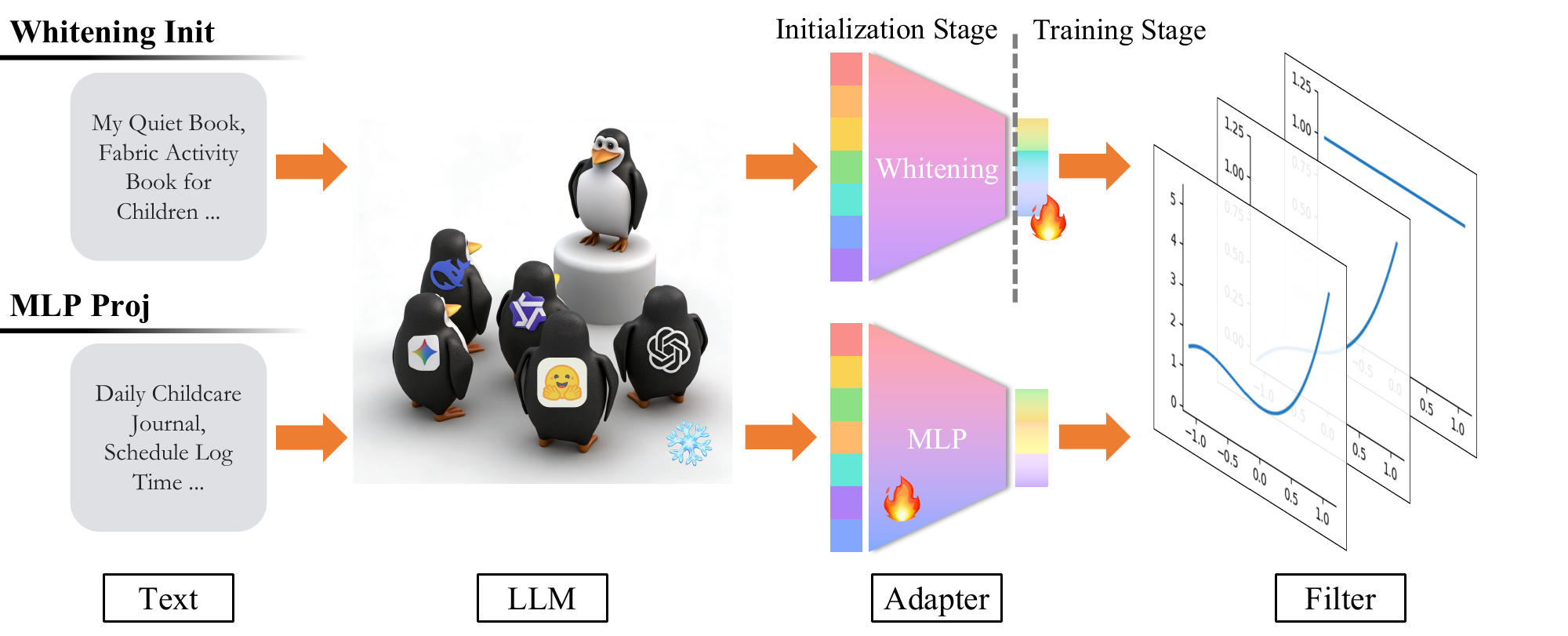}} \hfill
      \subfloat[(b) Case Study]{\includegraphics[width=0.3\textwidth]{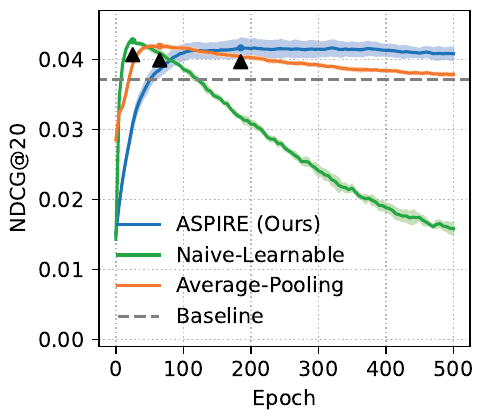}} \hfill
      
  \caption{Two benchmarked LLM-assisted architectures and metric stability analysis of MiniLM-L6 with Whitening Init adapter.}
  \label{figure-llm-cf}
\end{figure}

\begin{table*}[t]
\centering

\caption{Performance comparison of LLM-assisted CF on Baby. Best results are in \textbf{bold}. Refer to Appendix~\ref{appendix-llm} for details of the LMs and supplementary results.}

\begin{threeparttable}
\scalebox{0.825}{
\begin{tabular}{cccccccccccc}
\toprule
\multirow{2}{*}[-0.5ex]{Adapter} & \multirow{2}{*}[-0.5ex]{Filter} 
& \multirow{2}{*}[-0.5ex]{\makecell{Metric\\Sta.}}
& \multirow{2}{*}[-0.5ex]{\makecell{Filter\\Sta.}} 
& \multirow{2}{*}[-0.5ex]{\makecell{Filter\\Learn.}}
& \multicolumn{2}{c}{MiniLM-L6}
& \multicolumn{2}{c}{Qwen2.5-7B} 
& \multicolumn{2}{c}{SFR-Emb} \\
\cmidrule(lr){6-7} \cmidrule(lr){8-9} \cmidrule(lr){10-11}
& & & & & R@20 & N@20 & R@20 & N@20 & R@20 & N@20 \\
\midrule

\multirow{3}{*}{Whitening Init}
& Avg-P   & \xmark & \cmark & \xmark & 0.0928 & 0.0407 & 0.0931 & 0.0407 & 0.0961 & 0.0424 \\
& Naive-L & \xmark & \xmark     & \cmark & 0.0927 & 0.0407 & 0.0929 & 0.0403 & 0.0968 & \textbf{0.0430} \\
& ASPIRE  & \cmark & \cmark & \cmark & \textbf{0.0949} & \textbf{0.0415} & \textbf{0.0952} & \textbf{0.0418} & \textbf{0.0980} & \textbf{0.0430} \\

\midrule

\multirow{3}{*}{MLP Proj}
& Avg-P   & \cmark & \cmark & \xmark & 0.0831 & 0.0360 & 0.0859 & 0.0374 & 0.0889 & 0.0390 \\
& Naive-L & \xmark     & \xmark     & \cmark & 0.0755 & 0.0325 & 0.0775 & 0.0333 & 0.0801 & 0.0350 \\
& ASPIRE  & \cmark & \cmark & \cmark & \textbf{0.0935} & \textbf{0.0409} & \textbf{0.0928} & \textbf{0.0404} & \textbf{0.0926} & \textbf{0.0404} \\

\bottomrule
\end{tabular}
}

\begin{tablenotes}
\begin{minipage}{0.95\textwidth}
\footnotesize
\item[1] Sta.~=~Stability, Learn.~=~Learnability.
\end{minipage}
\end{tablenotes}

\end{threeparttable}
\label{table-llm-cf}
\end{table*}

Large Language Models (LLMs) have been widely adopted in recommender systems. 
Unlike Section~\ref{section-experiment}, which adopts multimodal information to construct graph structures, this section investigates the effectiveness of ASPIRE when employing LLMs as encoders to deliver powerful semantic representations for CF.
We consider two representative adapter designs, as shown in Figure~\ref{figure-llm-cf}(a).
Table~\ref{table-llm-cf} summarizes the overall performance, while Figure~\ref{figure-llm-cf}(b) illustrates metric stability for a representative case.\footnote{The relative performance between Table~\ref{table-llm-cf} (test results) and Figure~\ref{figure-llm-cf}(b) (validation trajectories) may differ slightly, potentially due to overfitting, especially for methods exhibiting unstable training dynamics.} 

In \textbf{Whitening Initialization} setting, we reduce the dimensionality of LLM-encoded features to the embedding space via whitening~\citep{su2021whitening, huang2021whitening}, which is proved to be effective in recommendation~\citep{zhang2024whitenrec, zhang2024dwsrec, shi2025recxplore}.
We then use the resulting representations as initialization for subsequent training~\citep{zhang2025llminit}.
This design is inspired by STAIR~\citep{xu2024stair}, which introduces step-wise convolution (a tunable graph filter) to mitigate modality erasure across deep convolutional layers.
In contrast, ASPIRE inherently alleviates this issue by enabling adaptive spectral approximation. 
Interestingly, Avg-P does not exhibit full metric stability in this setting. 
We observe that its performance gradually deteriorates after around 65 epochs, eventually degrading to the NDCG level achieved by randomly initialized Avg-P. 
We hypothesize that the degradation is caused by modality forgetting, as discussed in STAIR. 
Encouragingly, this phenomenon is significantly mitigated by ASPIRE.
In addition, we observe that Naive-L achieves relatively strong performance in this scenario, in contrast to Table~\ref{table-metric-comparison-main}, although low-frequency explosion still persists.
We attribute this to the high-quality initialization provided by LLM features, where amplifying low-frequency components in the early stages can already yield competitive predictions.
Overall, ASPIRE consistently achieves both metric and filter stability.

In \textbf{MLP Projection} setting, we freeze the LLM encoder and project the extracted features into the embedding space using an MLP~\citep{shi2025recxplore}. 
User representations are constructed by applying average pooling over item features, followed by MLP projection.
This design is inspired by AlphaRec~\citep{sheng2024alpharec}, which adopts a fixed Avg-P filter during the filtering stage. 
Despite the substantial change in embedding strategy, we observe that the stability patterns across different language models remain consistent with previous observations in this study.
Without strong initialization, Naive-L degrades into poor performance. 
These results highlight the importance of stable and adaptive filter learning in LLM-assisted CF, with ASPIRE offering a robust and generalizable solution.

\section{Related Work}



\paragraph{Spectral Graph Collaborative Filtering.}

Spectral methods are widely used in CF due to their interpretability and powerful graph signal processing techniques.
LightGCN~\citep{he2020lightgcn} simplifies NGCF~\citep{wang2019ngcf}, and can be interpreted as a low-pass graph filter.
More recently, ~\citet{guo2023jgcf} proposed a spectral analysis paradigm for CF and adopted Jacobi-basis for filtering; ~\citet{liu2025simgcf} measured signal contributions by comparing model performance under different filter settings.
These methods lack generalizability, as they are either restricted to specific architectures or burdened by high computational costs, thus yielding only heuristic filters that require cumbersome hyperparameter tuning.
Moreover, as emphasized by \citet{he2025ssc}, directly deploying advanced learnable spectral GNN architectures~\citep{wang2022jacobiconv, hordan2025equiepnn} to CF tasks remains infeasible.
Therefore, there is an urgent demand to deeply investigate the inherent failure causes and develop a graph filter adaptive to diverse graph structures and signal distributions.

\paragraph{Collaborative Filtering with Side Information.}

Existing side-information-based CF methods typically employ empirical graph filters, such as L-$n$~\citep{zhang2021lattice, ong2025smore} and Avg-P~\citep{lin2024gume, zhou2025cm3}, while recent approaches explore more flexible spectral designs~\citep{xu2024stair, he2025ssc}.
However, these methods still rely heavily on manually designed filter forms and task-specific hyperparameter tuning, whereas ASPIRE learns adaptive filters under a unified configuration.
Detailed discussions are provided in Appendix~\ref{appendix-related-work}.

\paragraph{Bi-level Optimization in Recommendation.}

Bi-level optimization (BLO) has been introduced into recommendation to model nested dependencies between primary objectives and auxiliary optimization processes~\citep{colson2007blooverview, wu2026unibio}.
Existing studies mainly employ BLO for specific downstream purposes, including feedback denoising~\citep{wang2023bod}, tokenizer-recommender alignment~\citep{bai2025bloger}, etc.
Different from these applications, we leverage BLO to mitigate the bias inherent in conventional optimization objectives.
A broader review is provided in Appendix~\ref{appendix-related-work}.

\section{Conclusion}

We propose ASPIRE, an adaptive spectral graph collaborative filtering framework that effectively addresses the low-frequency explosion issue and does not rely on prior knowledge of the underlying graph spectrum.
It achieves performance comparable to heuristic-based and heavily hyperparameter-dependent methods, while exhibiting stable training behavior.

\bibliography{refs}
\bibliographystyle{iclr2027/iclr2027_conference}

\newpage
\appendix
\tableofcontents

\newpage

\section{Additional Related Work}
\label{appendix-related-work}

\paragraph{Collaborative Filtering with Side Information.}

In CF, Side information is usually modeled as graph structures and processed by graph encoders.
Existing methods typically rely on empirical filters, such as L-$n$~\citep{zhang2021lattice, guo2024lgmrec, ong2025smore} and Avg-P~\citep{jiang2024share, lin2024gume, zhou2025cm3}. 
FREEDOM~\citep{zhou2023freedom} introduces a concise dual-branch framework but overlooks the inherent spectral characteristics of this scenario. 
STAIR~\citep{xu2024stair} focuses on the modality erasure issue and expands the search space of graph filter via step-wise convolution. 
SSC~\citep{he2025ssc} extends spectral filter design to heterogeneous graphs and devises a correction module to alleviate spectrum shifts. 
Despite improved spectral-domain flexibility, STAIR and SSC remains experiment-driven with non-unified hyperparameter settings, leading to substantial tuning overhead and poor generalization.
In contrast, the adaptive filter proposed in this paper achieves unified hyperparameter configuration, superior spectral adaptivity and stability. 

\paragraph{Bi-level Optimization in Recommendation.}

Bi-level optimization (BLO)~\citep{colson2007blooverview, wu2026unibio} can address nested interdependencies between core recommendation objectives and auxiliary tasks through nested optimization. 
For instance, to denoise implicit feedback, BOD~\citep{wang2023bod} proposes a gradient-matching method for the upper-level optimization to dynamically adjust sample weights. 
For generative recommendation, BLOGER~\citep{bai2025bloger} reformulates the task into a BLO framework to bridge the misalignment between the tokenizer and the recommender. 
Other application scopes of BLO include promoting fairness~\citep{zhang2025bifair}, providing explanations~\citep{wu2025binas}, facilitating cross-domain capability~\citep{chen2023biao}, etc~\citep{yin2024dr4sr}. 
In this work, we leverage BLO to avoid the bias inherent in conventional objectives.

\section{Proofs}
\label{appendix-proof}

\subsection{Proof of Theorem~\ref{theorem-loss-bound}}
\label{appendix-proof-bpr}

Under the Perfect Uniformity assumption defined by~\citet{wang2022directau}, for a nearly optimized model, the representation $h_u$ and $h_i$ uniformly distribute on the surface of the $d$-dimensional unit sphere.
Consider dot product as similarity function of two nodes $s(a, b) = h_a^\top h_b \in [-1, 1]$.
Thus we have
\begin{equation}
\label{eq-posneg-relation}
\begin{aligned}
0 &= |\mathcal{N}|^2\mathbb{E}_{n_1, n_2\sim P_\mathcal{N}}\left[s(n_1, n_2)\right] \\
&= \rho |\mathcal{U}||\mathcal{I}|\mathbb{E}_{(u, i^+)\sim P_{\mathrm{pos}}} \left[s(u, i^+)\right] 
+ (1-\rho) |\mathcal{U}||\mathcal{I}|\mathbb{E}_{(u, i^-)\sim P_{\mathrm{neg}}} \left[s(u, i^-)\right] \\
&\,\,\,\,\,\,\, + |\mathcal{U}|^2 \mathbb{E}_{u_1, u_2\sim P_\mathcal{U}} \left[s(u_1, u_2)\right]  
+ |\mathcal{I}|^2 \mathbb{E}_{i_1, i_2\sim P_\mathcal{I}} \left[s(i_1, i_2)\right]  \\
&= \rho |\mathcal{U}||\mathcal{I}|\mathbb{E}_{(u, i^+)\sim P_{\mathrm{pos}}}\left[s(u, i^+)\right] + (1-\rho) |\mathcal{U}||\mathcal{I}|\mathbb{E}_{(u, i^-)\sim P_{\mathrm{neg}}}\left[s(u, i^-)\right],
\end{aligned}
\end{equation}
where \(P_\mathcal{N}\), \(P_\mathcal{U}\), and \(P_\mathcal{I}\) are the uniform distributions over \(\mathcal{N}\), \(\mathcal{U}\), and \(\mathcal{I}\), respectively, while \(P_{\mathrm{pos}}\) and \(P_{\mathrm{neg}}\) denote the uniform distributions over the positive and negative user-item pairs, respectively. 
Moreover, \(\rho = \frac{|\mathcal{E}|}{|\mathcal{U}||\mathcal{I}|} \in (0,1)\) denotes the edge density of the dataset.

We denote 
\[
s^+ = s(u, i^+),\quad s^- = s(u, i^-),\quad t = 1-\sigma(s^+ - s^-) \in \left[\frac{1}{e^2+1}, \frac{e^2}{e^2+1}\right],
\]
\[
\nabla^+ = \frac{\partial s^+}{\partial g(\lambda_k)},\quad \nabla^- = \frac{\partial s^-}{\partial g(\lambda_k)},\quad \nabla = \nabla^+ - \nabla^-.
\]
Taking the partial derivative of both sides of ~\eqref{eq-posneg-relation} with respect to $g(\lambda_k)$, we obtain
\[
\rho \mathbb{E}\left[ \nabla^+ \right] + (1-\rho) \mathbb{E}\left[ \nabla^- \right]
= \rho \frac{\partial \mathbb{E}\left[s^+\right]}{\partial g(\lambda_k)} + (1-\rho)\frac{\partial \mathbb{E}\left[s^-\right]}{\partial g(\lambda_k)}
= 0
\]
So we have
\[
\mathbb{E}\left[\nabla\right] = \mathbb{E}\left[\nabla^+\right] - \mathbb{E}\left[\nabla^-\right] = \frac{1}{1-\rho} \mathbb{E}\left[\nabla^+\right].
\]

The BPR loss is defined as
\[
\ell_{\mathrm{BPR}} = -\mathbb{E}_{(u, i^+)\sim P_{\mathrm{pos}}} \mathbb{E}_{(u, i^-)\sim P_{\mathrm{neg}}} \log\left( \sigma(s^+ - s^-)\right).
\]
The derivative of the spectral response is
\[
\begin{aligned}
\frac{\partial \ell_{\mathrm{BPR}}}{\partial g(\lambda_k)} 
&= 
-\mathbb{E}\left[(1-\sigma(s^+ - s^-))\left(\frac{\partial s^+}{\partial g(\lambda_k)}-\frac{\partial s^-}{\partial g(\lambda_k)}\right)\right] \\
&= -\mathbb{E}[t\nabla] \\
&= - \mathbb{E}[t] \mathbb{E}\left[\nabla\right] -\mathbb{E}[t(\nabla - \mathbb{E}\left[\nabla\right])] \\
&= -\frac{1}{1-\rho}\mathbb{E}[t] \mathbb{E}\left[\nabla^+\right] + \epsilon
\end{aligned}
\]
where
\begin{equation}
|\epsilon| < \delta = \mathbb{E}\left[|\nabla - \mathbb{E}\left[\nabla\right]|\right] \approx 0.
\label{equation-negligible-delta}
\end{equation}

Next we compute $\mathbb{E}\left[\nabla^+\right]$.

Let $w_{ui}=P_{\mathrm{pos}}(u,i)$ be the sampling probability of the positive pair $(u,i)$ from $A$, and $H\in\mathbb{R}^{|\nodes|\times d}$ be the matrix stacking all node embeddings. Then

\[
\mathbb{E}_{(u,i)\sim P_{\mathrm{pos}}}[s(u, i)]
= \sum_{u,i} w_{ui} \, h_u^\top h_i
= \mathrm{Tr}(H^\top W H),
\quad
W = [w_{ui}],
\quad
\sum_{u,i} w_{ui} = 1 .
\]

Under uniform edge sampling, we have $W = \frac{1}{2|\mathcal{E}|} A = \phi(A)$. 
Further substitute the spectral filtering form $H = U g(\Lambda) U^\top E$, we obtain
\[
\begin{aligned}
\mathbb{E}[s^+]
&=
\mathrm{Tr}\!\left(
E^\top U g(\Lambda) U^\top \phi(A) \, U g(\Lambda) U^\top E
\right) \\
&=
\mathrm{Tr}\!\left(
E^\top U g(\Lambda) U^\top U \Lambda U^\top \, U g(\Lambda) U^\top E
\right) \\
&=
\mathrm{Tr}\!\left(
E^\top U g(\Lambda) \Lambda g(\Lambda) U^\top E
\right) \\
&=
\mathrm{Tr}\!\left(
g(\Lambda) \Lambda g(\Lambda) \, U^\top E E^\top U
\right)
\end{aligned}
\]

Let $\tilde{E} = U^\top E$, then
\begin{equation}
\label{equation-posneg-filter}
\begin{aligned}
\mathbb{E}[s^+]
&=
\mathrm{Tr}\!\left(
g(\Lambda) \Lambda g(\Lambda) \, \tilde{E}\tilde{E}^\top
\right) \\
&=
\sum_k
\lambda_k g(\lambda_k)^2
\left(\tilde{E}\tilde{E}^\top\right)_{kk} \\
&=
\sum_k
\lambda_k g(\lambda_k)^2
\|
\tilde{E}_k
\|^2.
\end{aligned}  
\end{equation}

where $\tilde{E}_k$ is the $k$-th row of $\tilde{E}$.

Taking the partial derivative with respect to $g(\lambda_k)$ yields
\[
\mathbb{E}\left[\nabla^+\right]
=
\frac{\partial \mathbb{E}\left[s^+\right]}{\partial g(\lambda_k)}
=
2\lambda_k g(\lambda_k) \|\tilde{E}_k\|^2,
\]

Thus for $\lambda_k g(\lambda_k)\|\tilde{E}_k\|^2 > 5(1-\rho)\delta$,
\[
\begin{aligned}
\frac{\partial \ell_{\mathrm{BPR}}}{\partial g(\lambda_k)}
&< - \frac{2}{1-\rho} \lambda_k g(\lambda_k)\|\tilde{E}_k\|^2 \mathbb{E}[t] +\delta \\
&\le - \frac{2}{1-\rho} \lambda_k g(\lambda_k)\|\tilde{E}_k\|^2 \frac{1}{e^2+1} +\delta \\
&< - \frac{1}{5(1-\rho)} \lambda_k g(\lambda_k)\|\tilde{E}_k\|^2 + \delta < 0
\end{aligned}
\]
And for $\lambda_k g(\lambda_k)\|\tilde{E}_k\|^2 < -5(1-\rho)\delta$,
\[
\begin{aligned}
\frac{\partial \ell_{\mathrm{BPR}}}{\partial g(\lambda_k)}
&> - \frac{2}{1-\rho} \lambda_k g(\lambda_k)\|\tilde{E}_k\|^2 \mathbb{E}[t] -\delta \\
&\ge - \frac{2}{1-\rho} \lambda_k g(\lambda_k)\|\tilde{E}_k\|^2 \frac{1}{e^2+1} -\delta \\
&> - \frac{1}{5(1-\rho)} \lambda_k g(\lambda_k)\|\tilde{E}_k\|^2 - \delta > 0
\end{aligned}
\]
This completes the proof.

\paragraph{Empirical Verification of the Perfect Uniformity Assumption.}

\begin{figure}
  \centering
  \includegraphics[width=0.99\textwidth]{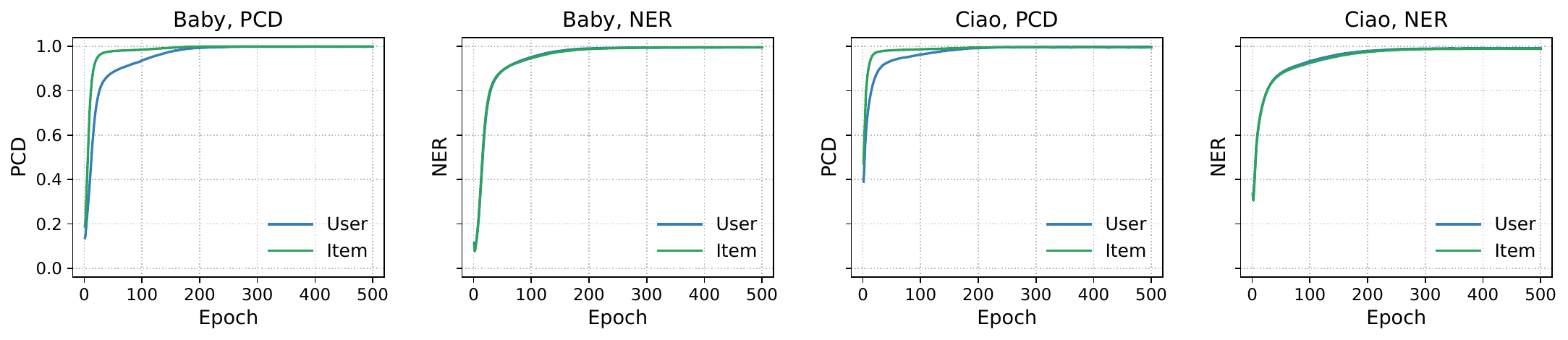}
  \caption{Evolution of uniformity metrics during training.}
  \label{figure-exp-uniformity-metrics}
\end{figure}

To empirically assess the Perfect Uniformity assumption underlying our theorem, we track two uniformity metrics throughout training.
Pairwise Cosine Distance (PCD) measures the discrepancy between node pairs, with larger values approaching 1 indicating better uniformity. 
$$\mathrm{PCD}=1-\frac{1}{N(N-1)}\sum_{i\ne j}\frac{\mathbf{h}_i^\top\mathbf{h}_j}{\|\mathbf{h}_i\|_2\|\mathbf{h}_j\|_2}$$
Normalized Effective Rank (NER) evaluates the utilization of spatial dimensions, with larger values approaching 1 indicating better uniformity. 
$$\mathrm{NER}=\frac{1}{d}\exp\left(-\sum_{k=1}^{d}p_k\log p_k\right),\quad p_k=\frac{\sigma_k}{\sum_{j=1}^{d}\sigma_j}  $$
where $\sigma_k$ denotes the $k$-th singular value of $H$.
As shown in the Figure~\ref{figure-exp-uniformity-metrics}, both metrics rapidly converge toward 1.00, providing empirical support for the approximate validity of the Perfect Uniformity assumption during the early stages of training.

\subsection{Extension to Cross-Entropy Loss}
\label{appendix-proof-ce}

An analogous result holds for the CE loss.

With $M = |\mathcal{I}| - 1$, the CE loss is defined as
\[
\ell_{\text{CE}} = -\mathbb{E}_{(u, i^+)\sim P_{\mathrm{pos}}}  \log\left( 
\frac{\exp(s^+)}{\exp(s^+) + \sum_{m=1}^M \exp(s^-_m)}
\right).
\]

We denote
\[
p^+ = \frac{\exp(s^+)}{\exp(s^+) + \sum_{m=1}^M \exp(s^-_m)}, \;\;
p^-_m = \frac{\exp(s^-_m)}{\exp(s^+) + \sum_{m=1}^M \exp(s^-_m)} \;
\in \left[ \frac{1}{Me^2+1}, \frac{e^2}{e^2+M}\right].
\]

The derivative of the spectral response is
\[
\begin{aligned}
\frac{\partial \ell_{\mathrm{CE}}}{\partial g(\lambda_k)} 
&= 
-\mathbb{E}\left[(1-p^+)\nabla^+ - \sum_{m=1}^M p_m^-\nabla^-_m\right] \\
&=
-\mathbb{E}\left[\sum_{m=1}^M p_m^-\nabla^+ - \sum_{m=1}^M p_m^-\nabla^-_m\right] \\
&= 
-\sum_{m=1}^M \mathbb{E}_{(u, i^+)\sim P_{\mathrm{pos}}}\left[p_m^- \nabla_m \right] \\
&= 
-M \mathbb{E}_{(u, i^+)\sim P_{\mathrm{pos}},\,m\sim \mathrm{Uniform}\{1,M\}} \left[p_m^- \nabla_m \right] \\
& \approx 
-M \mathbb{E}_{(u, i^+)\sim P_{\mathrm{pos}}, \;(u, i^-)\sim P_{\mathrm{neg}}} \left[p^- \nabla \right] \\
&= 
- M \mathbb{E}\left[ p^- \right] \mathbb{E}\left[ \nabla \right] - M\mathbb{E}\left[p^-(\nabla - \mathbb{E}\left[\nabla\right])\right] \\
&= 
-\frac{M}{1-\rho}\mathbb{E}[p^-] \mathbb{E}\left[\nabla^+\right] + \epsilon^{\prime}
\end{aligned}
\]
where $|\epsilon^{\prime}| \le \delta^{\prime} = \frac{Me^2}{e^2+M} \mathbb{E}\left[|\nabla - \mathbb{E}\left[\nabla\right]|\right]$.

$\mathbb{E}\left[\nabla^+\right]$ is consistent with the previous section.

Thus for $\lambda_k g(\lambda_k)\|\tilde{E}_k\|^2 > 5(1-\rho)\delta^{\prime}$,
\[
\begin{aligned}
\frac{\partial \ell_{\mathrm{CE}}}{\partial g(\lambda_k)}
&< - \frac{2}{1-\rho} \lambda_k g(\lambda_k)\|\tilde{E}_k\|^2 M\mathbb{E}[t] +\delta^{\prime} \\
&\le - \frac{2}{1-\rho} \lambda_k g(\lambda_k)\|\tilde{E}_k\|^2 \frac{M}{Me^2+1} +\delta^{\prime} \\
&< - \frac{1}{5(1-\rho)} \lambda_k g(\lambda_k)\|\tilde{E}_k\|^2 + \delta^{\prime} \\
& < 0 \quad \text{when } \delta^{\prime} \text{ is sufficiently small}
\end{aligned}
\]
And for $\lambda_k g(\lambda_k)\|\tilde{E}_k\|^2 < -5(1-\rho)\delta^{\prime}$,
\[
\begin{aligned}
\frac{\partial \ell_{\mathrm{CE}}}{\partial g(\lambda_k)}
&> - \frac{2}{1-\rho} \lambda_k g(\lambda_k)\|\tilde{E}_k\|^2 M\mathbb{E}[t] -\delta^{\prime} \\
&\ge - \frac{2}{1-\rho} \lambda_k g(\lambda_k)\|\tilde{E}_k\|^2 \frac{M}{Me^2+1} -\delta^{\prime} \\
&> - \frac{1}{5(1-\rho)} \lambda_k g(\lambda_k)\|\tilde{E}_k\|^2 - \delta^{\prime} \\
& > 0 \quad \text{when } \delta^{\prime} \text{ is sufficiently small}
\end{aligned}
\]

However, since $\delta' = \frac{M e^2}{e^2 + M},\delta$, the coefficient $\frac{M e^2}{e^2 + M}$ is monotonically increasing in $M$ and converges to $e^2$ as $M \to \infty$. 
Therefore, the sufficient condition here becomes slightly more restrictive as $M$ grows.

\section{Experimental Setups}
\label{appendix-exp-setup}

\subsection{Datasets}

\begin{table}[]
  \centering
  \caption{
    Dataset statistics.
    `\#U-I': the number of interactions.
    `\#I-I' the number of item-item edges.
    `\#U-U': the number of user-user edges.
  }
  \label{table-dataset}
\scalebox{1.}{
    \begin{tabular}{l *{6}{w{c}{1.3cm}}}
      \toprule
                & \#Users & \#Items & \#U-I     & \#I-I      & \#U-U     & Sparsity  \\
      \midrule
    Baby        & 19,445  & 7,050   & 160,792   & 141,000    & -    & 99.88\% \\
    Electronics & 192,403 & 63,001  & 1,689,188 & 1,260,020  & -    & 99.99\% \\
      \midrule
    Ciao        & 6,804   & 17,660  & 139,112   & -   & 106,433   & 99.88\% \\
    LastFM      & 1,875   & 4,613   & 77,465    & -   & 25,182    & 99.10\% \\
      \bottomrule
    \end{tabular}
}
\end{table}

The two multimodal datasets, Amazon Baby and Electronics\footnote[1]{https://github.com/enoche/MMRec}, were originally published by~\citet{he2016ups}.
Item title, brand, categories and description are encoded via all-MiniLM-L6-v2~\citep{wang2020minilm} as textual features.
Item thumbnails are encoded with CNN as visual features.
Among the two social datasets, Ciao\footnote[2]{https://www.cse.msu.edu/{\textasciitilde}tangjili/datasetcode/truststudy.htm} was released by~\citet{tang2012ciao1}, while LastFM\footnote[3]{https://github.com/Sherry-XLL/Social-Datasets/tree/main/lastfm} was provided by~\citet{cantador2011lastfm}.
The dataset statistics are presented in Table~\ref{table-dataset}.

\subsection{Baselines}
\label{appendix-baseline}

\begin{table}[]
  \centering
  \caption{
    Typical models using baseline filters.
  }
  \label{table-filter-examples}
  
\scalebox{0.72}{
\renewcommand{\arraystretch}{1.2}
\begin{tabular}{c|c|c|c}
    \toprule
    Filter & Tunable & Learnable & Models \\
    \midrule
    L-$0$ & \xmark & \xmark & MF~\citep{rendle2012bpr} \\
    L-$n$ & \cmark & \xmark & FREEDOM~\citep{zhou2023freedom}, LGMRec~\citep{guo2024lgmrec}, SMORE~\citep{ong2025smore} \\
    Avg-P & \cmark & \xmark & LightGCN~\citep{he2020lightgcn}, BM3~\citep{Zhou2023bm3}, SHaRe~\citep{jiang2024share}\\
    \midrule
    Jacobi & \cmark & \xmark & JGCF~\citep{guo2023jgcf}, SimGCF~\citep{liu2025simgcf} (a Jacobi-based variant) \\
    Lin-C & \cmark & \xmark & SSC~\citep{he2025ssc} \\
    \midrule
    Naive-L & \cmark & \cmark &  - \\
    \bottomrule
\end{tabular}
}
\end{table}

We present typical models for each baseline filter in Table~\ref{table-filter-examples}.






\subsection{Implementation Details}
\label{appendix-implementation}

We fix the embedding dimension to 64 and the batch size to 2048.
For embedding optimization, we adopt the Adam optimizer~\citep{kingma2014adam}, while filter parameters are optimized using SGD~\citep{robbins1951sgd}.
The learning rate of Adam is tuned within $\{1\mathrm{e}{-4}, 5\mathrm{e}{-4}, 1\mathrm{e}{-3}, 5\mathrm{e}{-3}\}$, and that of SGD within $\{1\mathrm{e}{-2}, 5\mathrm{e}{-2}, 1\mathrm{e}{-1}, 5\mathrm{e}{-1}\}$.
Weight decay is carefully tuned for all baselines, while it is fixed to 0 for our method.
For graph filtering, we set the filter order $L=3$ by default, except for Layer-$n$ where $L$ is searched over $\{1,2,3\}$.
Other model-specific hyperparameters are tuned according to the settings recommended in the original papers.

For the lower-level objective $\mathcal{L}_{\mathrm{train}}$ of ASPIRE, we employ the standard BPR loss to ensure fair comparison with existing methods.
For the upper-level objective $\mathcal{L}_{\mathrm{valid}}$, we consider two loss variants: BPR and CE.
When using the CE loss, we introduce a temperature parameter $\tau$ and tune it over $\{0.5, 0.75, 1.0, 1.25, 1.5, 1.75, 2.0\}$.
Following our decoupled training strategy, we randomly split 50\% of the validation data as an auxiliary set for upper-level optimization, while the remaining 50\% is used for hyperparameter tuning.
We fix the upper-level update interval $T$ to 5 in our bi-level optimization scheme.

\section{Additional Experimental Results}

\subsection{Extended Stability Evaluation for Figure~\ref{fig-intro-group}}
\label{appendix-stability}

\paragraph{Metric Stability.}

\begin{figure}[]
  \centering
  \hspace*{0.0001\textwidth}
  \subfloat[(a) Baby, Heterogeneous]{\includegraphics[width=0.325\textwidth]{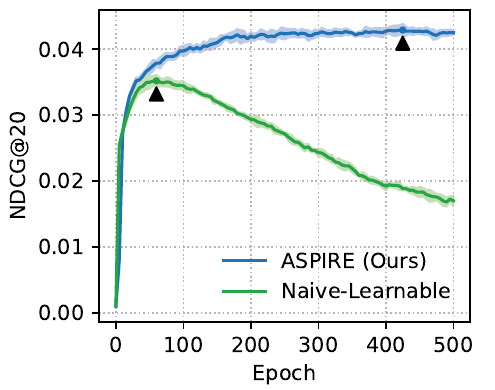}} \hfill
  \subfloat[(b) Ciao, Homogeneous]{\includegraphics[width=0.325\textwidth]{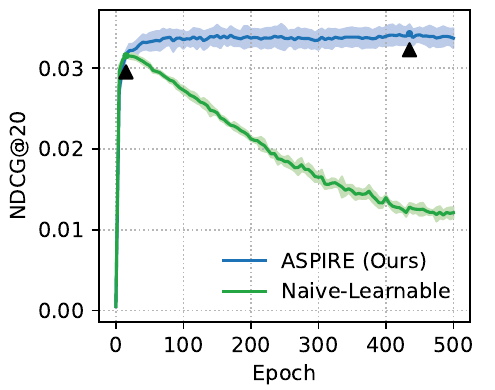}} \hfill
  \subfloat[(c) Ciao, Heterogeneous]{\includegraphics[width=0.325\textwidth]{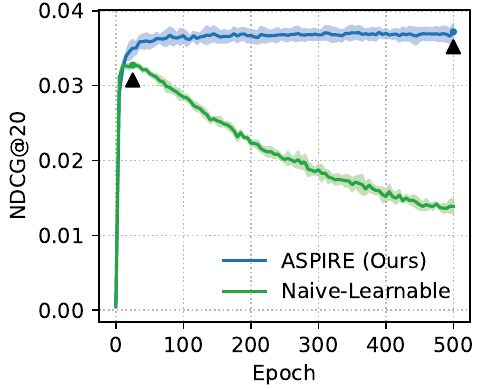}} 
  \hspace*{0.01\textwidth} 
\caption{
    NDCG@20 Trajectory of ASPIRE and Naive-L.
}
\label{fig-exp-addition-metric-stab}
\end{figure}

In Figure~\ref{fig-exp-addition-metric-stab}, we extend Figure~\ref{fig-intro-group} (a) to the Heterogeneous scenario and the Ciao dataset, demonstrating that the flaws of Naive-L in metric stability are universally prevalent.

\paragraph{Filter Stability.}

\begin{figure}
  \centering
  \includegraphics[width=0.75\textwidth]{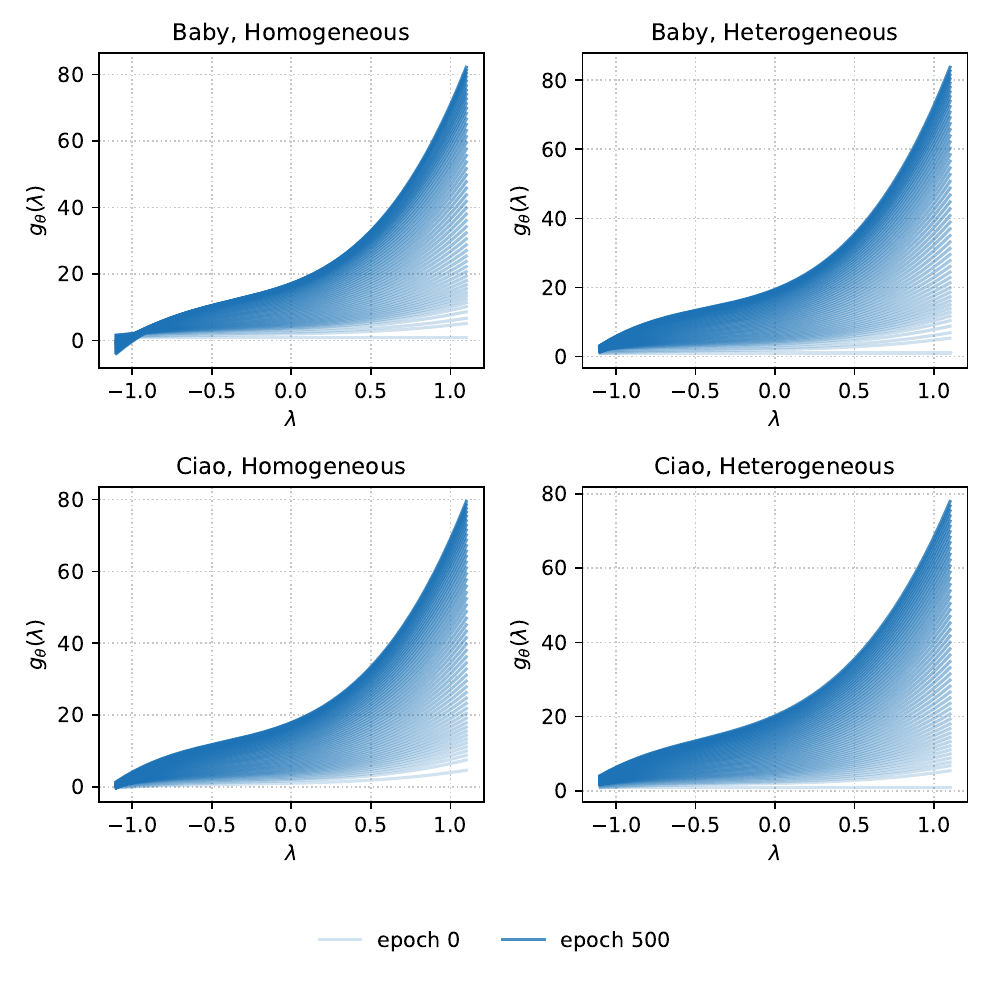}
  \caption{Filter evolution of Naive-L.}
  \label{figure-exp-filter-evolution-inone-naivel}
\end{figure}

\begin{figure}
  \centering
  \includegraphics[width=0.75\textwidth]{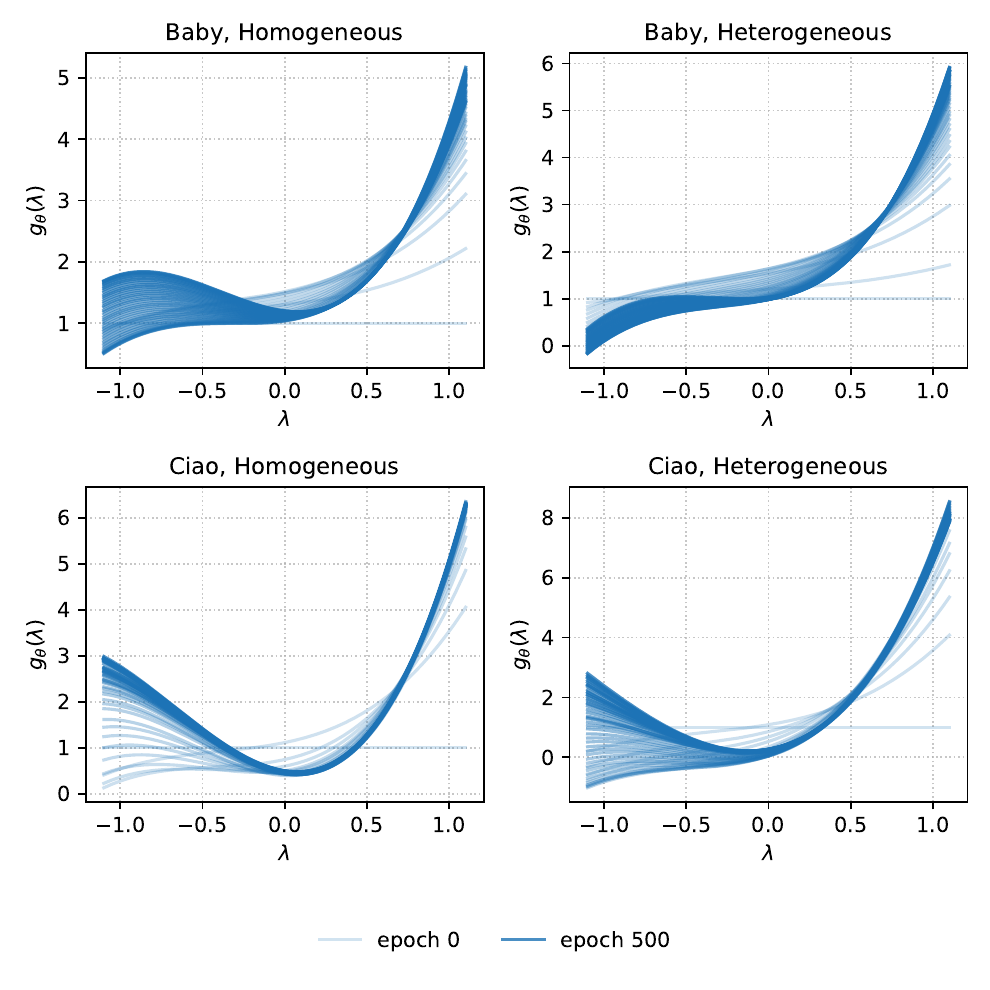}
  \caption{Filter evolution of ASPIRE.}
  \label{figure-exp-filter-evolution-inone-aspire}
\end{figure}


\begin{figure}
  \includegraphics[width=0.99\textwidth]{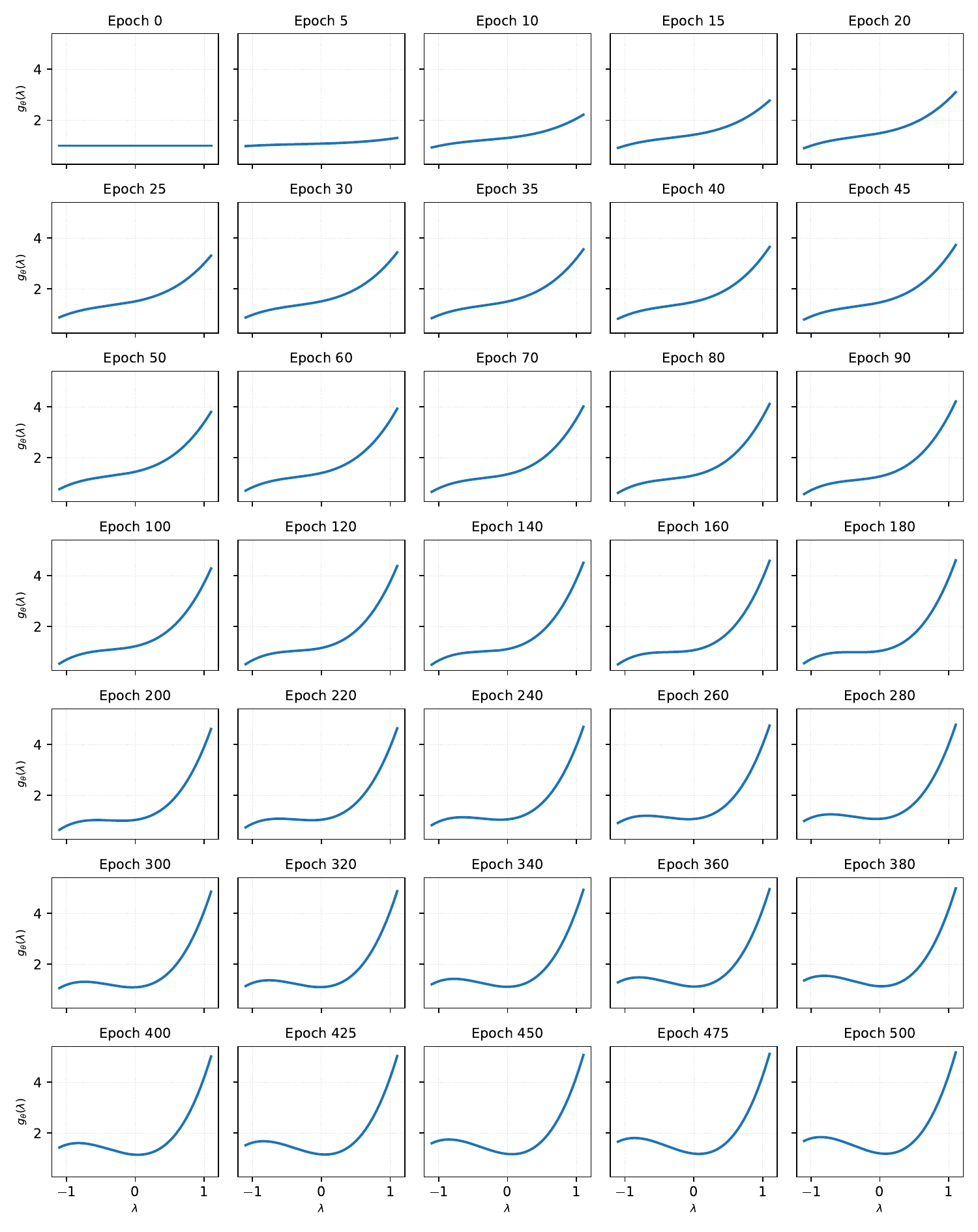}
  \caption{Filter evolution of ASPIRE under Homogeneous scenario on Baby.}
  \label{figure-exp-filter-evolution-aspire-baby}
\end{figure}


\begin{figure}
  \includegraphics[width=0.99\textwidth]{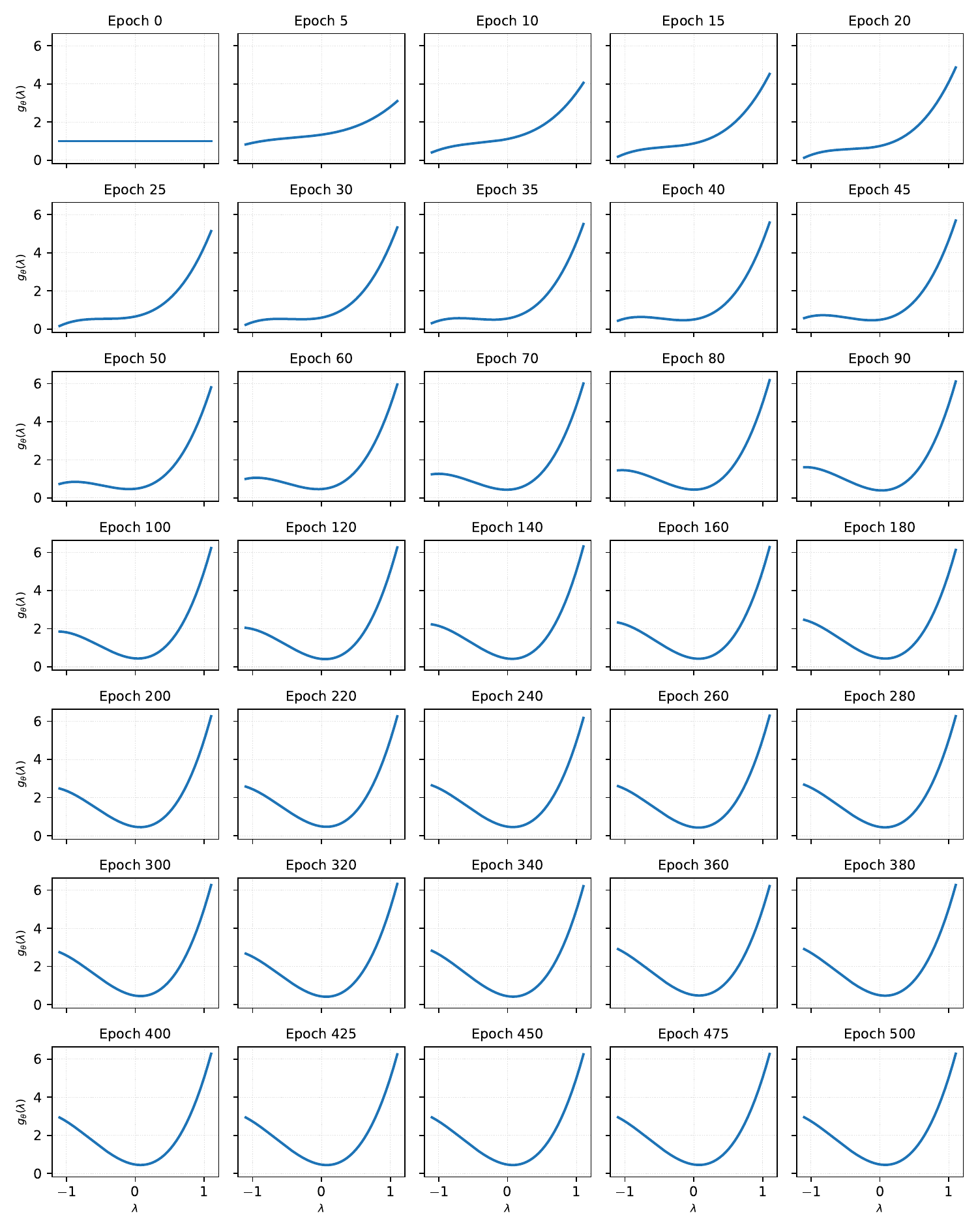}
  \caption{Filter evolution of ASPIRE under Homogeneous scenario on Ciao.}
  \label{figure-exp-filter-evolution-aspire-ciao}
\end{figure}

This section aims to clearly demonstrate how the filters evolve throughout the training process, serving as a more detailed version of Figure~\ref{fig-intro-group} (b) and (c).
In Figure~\ref{figure-exp-filter-evolution-inone-naivel} and Figure~\ref{figure-exp-filter-evolution-inone-aspire}, we plot one curve every 10 epochs with gradually increasing opacity to illustrate the evolution of the spectral response function.
To make the evolution of ASPIRE easier to observe, we further sample 35 curves across training and plot them separately, resulting in Figure~\ref{figure-exp-filter-evolution-aspire-baby} and Figure~\ref{figure-exp-filter-evolution-aspire-ciao}.

\subsection{Stability Evaluation for CE Loss}
\label{appendix-stab-ce}

\paragraph{Metric Stability.}

\begin{figure}[]
  \centering
  \hspace*{0.0001\textwidth}
  \subfloat[(a) Baby]{\includegraphics[width=0.45\textwidth]{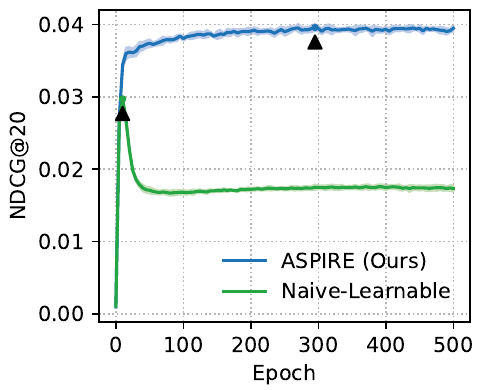}} \hfill
  \subfloat[(b) Ciao]{\includegraphics[width=0.45\textwidth]{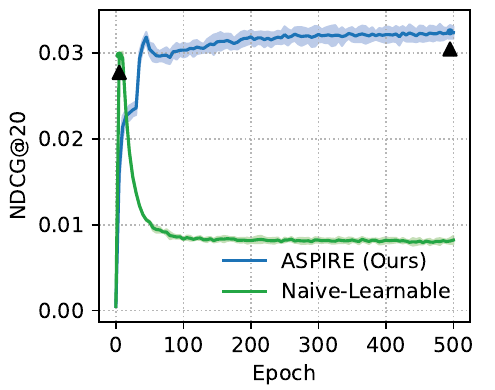}} \hfill
  \hspace*{0.0001\textwidth} 
\caption{
    NDCG@20 Trajectory of ASPIRE and Naive-L under CE loss.
}
\label{fig-exp-ce-metric-stab}
\end{figure}

As shown in Figure~\ref{fig-exp-ce-metric-stab}, Naive-L also suffers from a sudden performance drop when cross-entropy (CE) loss is adopted for filter learning.
Different from BPR loss, the model converges rapidly to an inferior performance plateau, which further demonstrates the inherent gap in supervision signals.

\paragraph{Filter Stability.}

\begin{figure}
  \centering
  \includegraphics[width=0.75\textwidth]{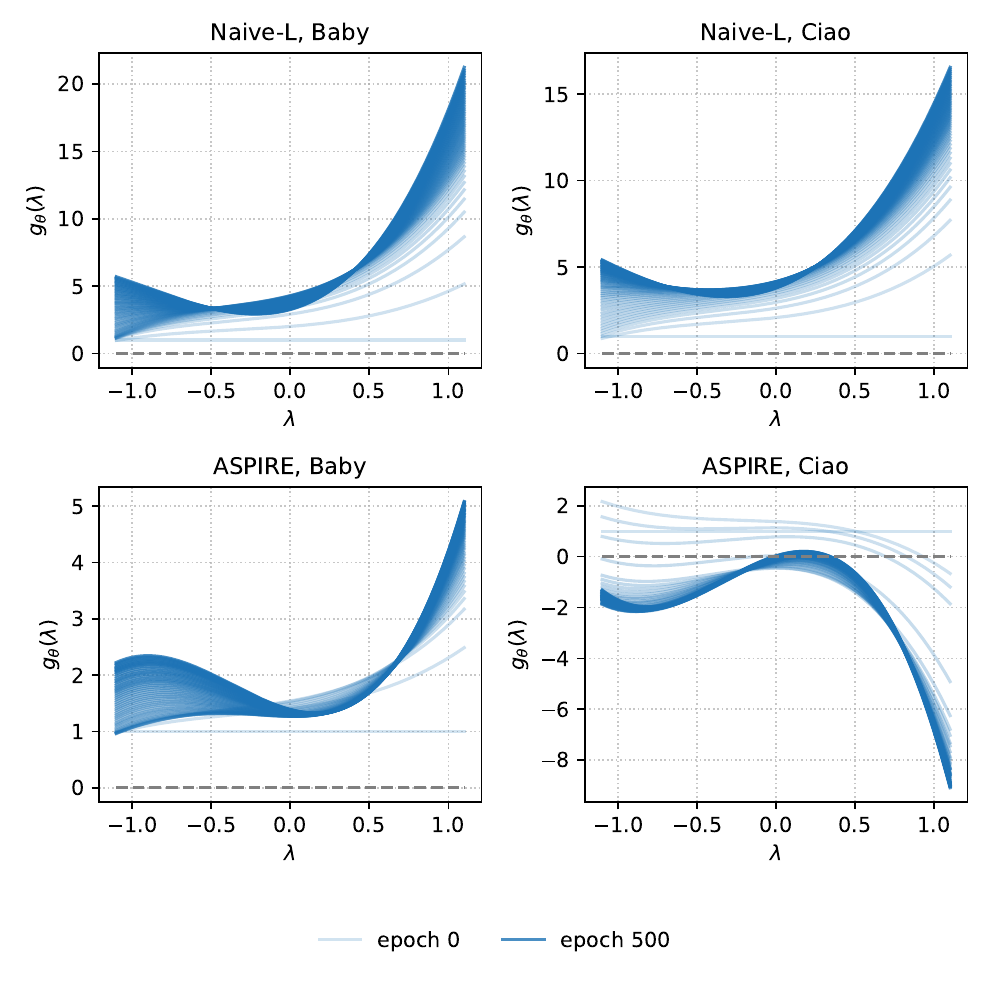}
  \caption{Filter evolution of Naive-L and ASPIRE under CE.}
  \label{figure-exp-ce-filter-stab}
\end{figure}

\begin{figure}
  \centering
  \includegraphics[width=0.8\textwidth]{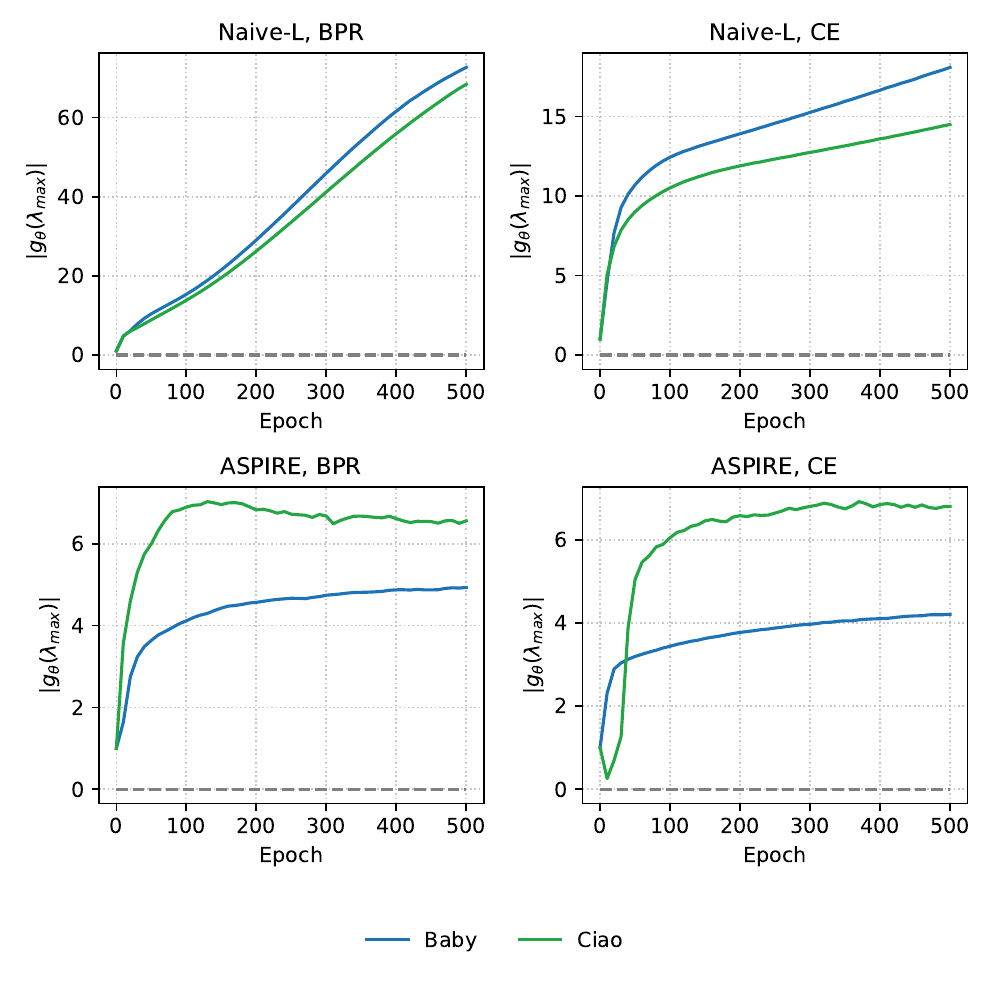}
  \caption{Low-frequency explosion from a quantitative perspective, derived from Figures~\ref{figure-exp-filter-evolution-inone-naivel},~\ref{figure-exp-filter-evolution-inone-aspire} and~\ref{figure-exp-ce-filter-stab}. In Homogeneous setting, $\lambda_{max}=1$.}
  \label{figure-exp-mixloss-maxeigenv}
\end{figure}

As shown in Figure~\ref{figure-exp-ce-filter-stab}, for Naive-L, although high-frequency components are not suppressed as strongly as under the BPR loss, the low-frequency explosion issue still persists, evidenced by the consistently large and increasing $g_\theta(\lambda_{max})$.
Since the convergence trend is not straightforward to observe from the above figure, we further visualize the evolution of $|g_\theta(\lambda_{max})|$ over the entire training process in Figure~\ref{figure-exp-mixloss-maxeigenv}.
For Naive-L, the magnitude of $|g_\theta(\lambda_{\text{max}})|$ spikes sharply within the initial roughly 20 epochs. 
It continues to rise rapidly under BPR optimization, while exhibiting a slower growth rate with the CE loss, and both approaches maintain an approximately linear upward trend.
Even with the milder growth brought by CE loss, its persistent upward trend remains substantial in contrast to the near-stagnant convergence behavior of ASPIRE, thereby proving that low-frequency explosion still prevails.
A possible reason for the discrepancy in their growth slopes is explained in Appendix~\ref{appendix-proof-ce}.

\subsection{Stability Evaluation for Table~\ref{table-llm-cf}}
\label{appendix-llm}

\paragraph{Metric Stability.}

\begin{figure}
  \includegraphics[width=0.99\textwidth]{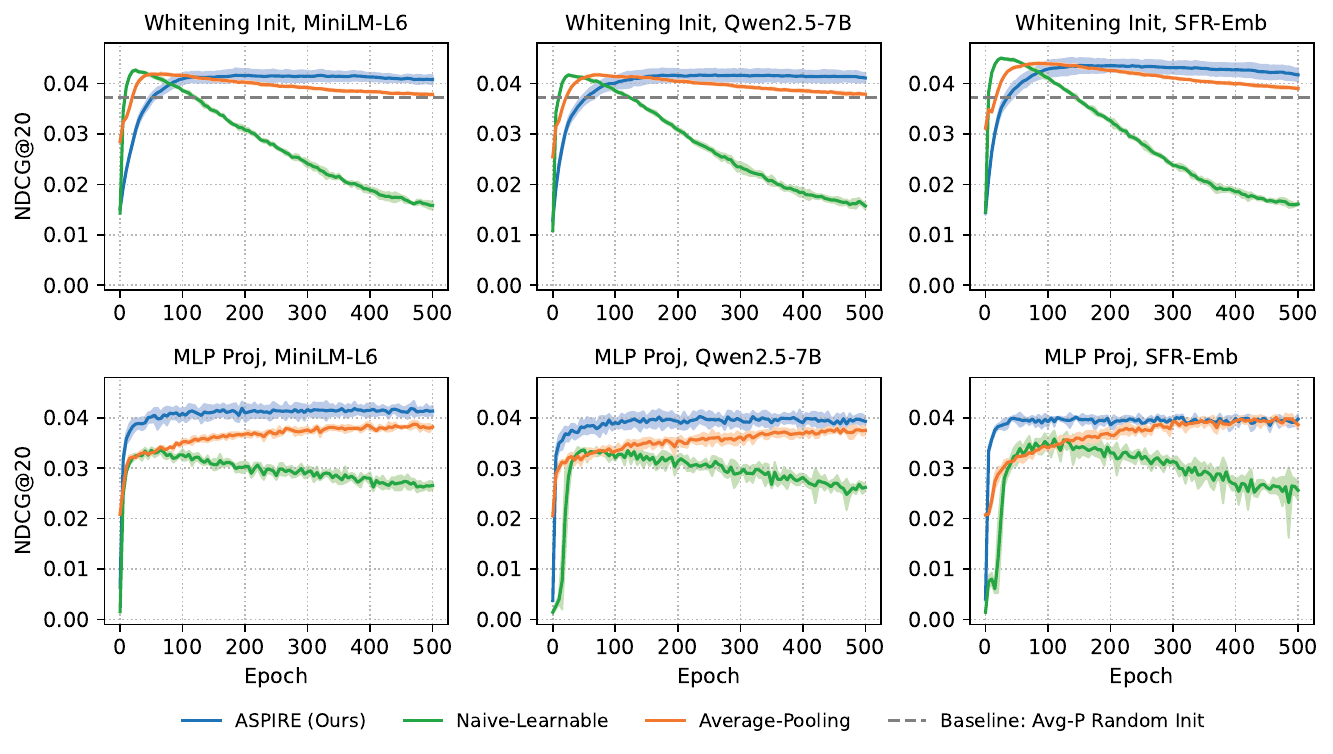}
  \caption{Metric Stability in LLM-powered CF.}
  \label{figure-exp-llm-metric-stab-mixed-baby}
\end{figure}

As shown in Figure~\ref{figure-exp-llm-metric-stab-mixed-baby}, ASPIRE maintains consistent stability throughout training, while Naive-I exhibits obvious instability due to an early performance decline in the training process.
Avg-P behaves unstably under the Whitening Init setting, with its performance gradually degrading toward the baseline, as elaborated in the main text.
In contrast, it remains stable under the MLP Proj setting.
It is also worth noting that Avg-P achieves better initial performance than the other two methods, and the Whitening Init setup outperforms the MLP Proj setup at initialization.
This phenomenon reflects the differences in initialization quality: low-pass filter-based initialization presents clear advantages over full-pass initialization, and direct adoption of semantic embeddings delivers better results than a randomly initialized MLP.

\paragraph{Filter Stability.}

\begin{figure}
  \includegraphics[width=0.99\textwidth]{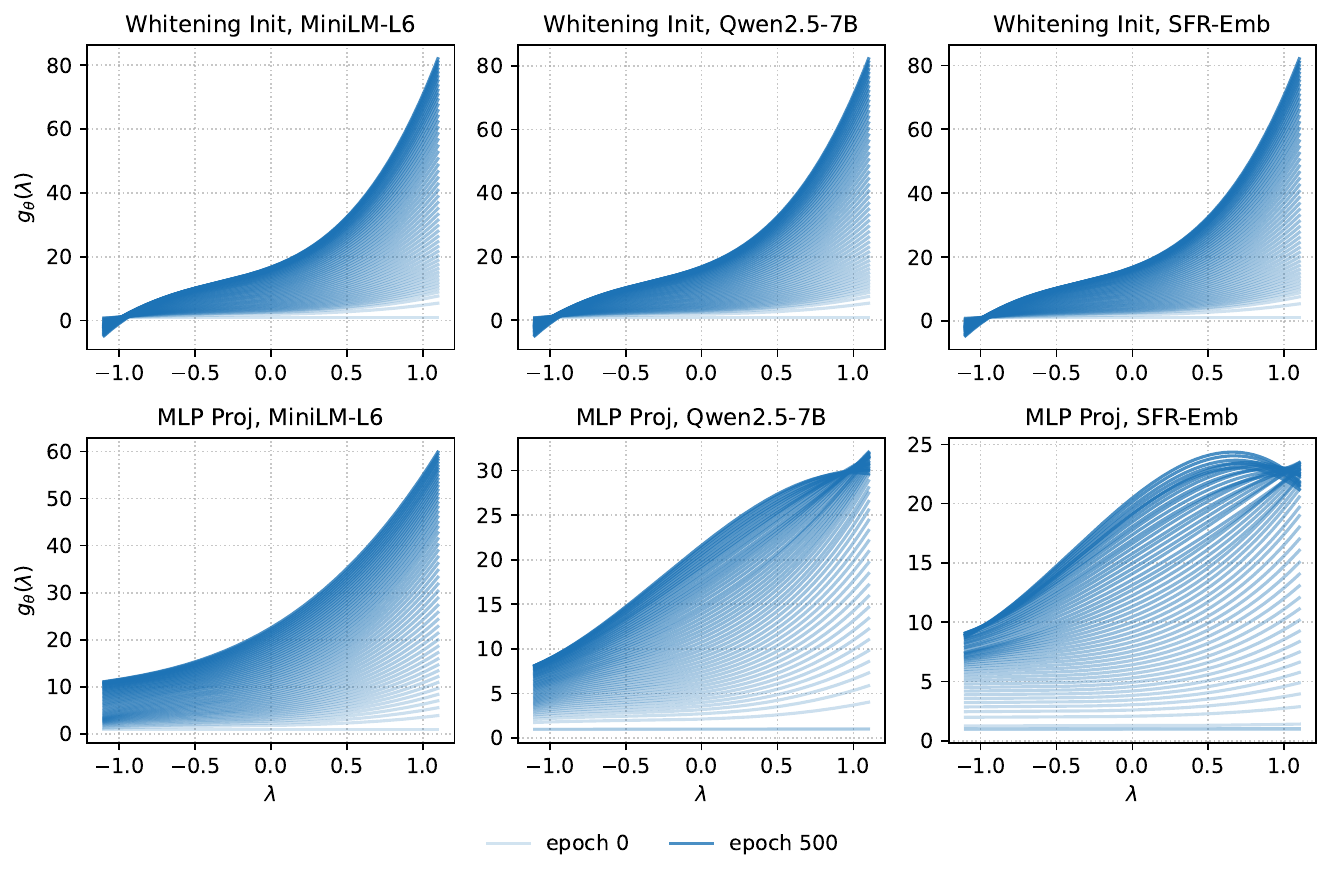}
  \caption{Filter Stability of Naive-L in LLM-powered CF.}
  \label{figure-exp-llm-filter-stab-naivel-baby}
\end{figure}

\begin{figure}
  \includegraphics[width=0.99\textwidth]{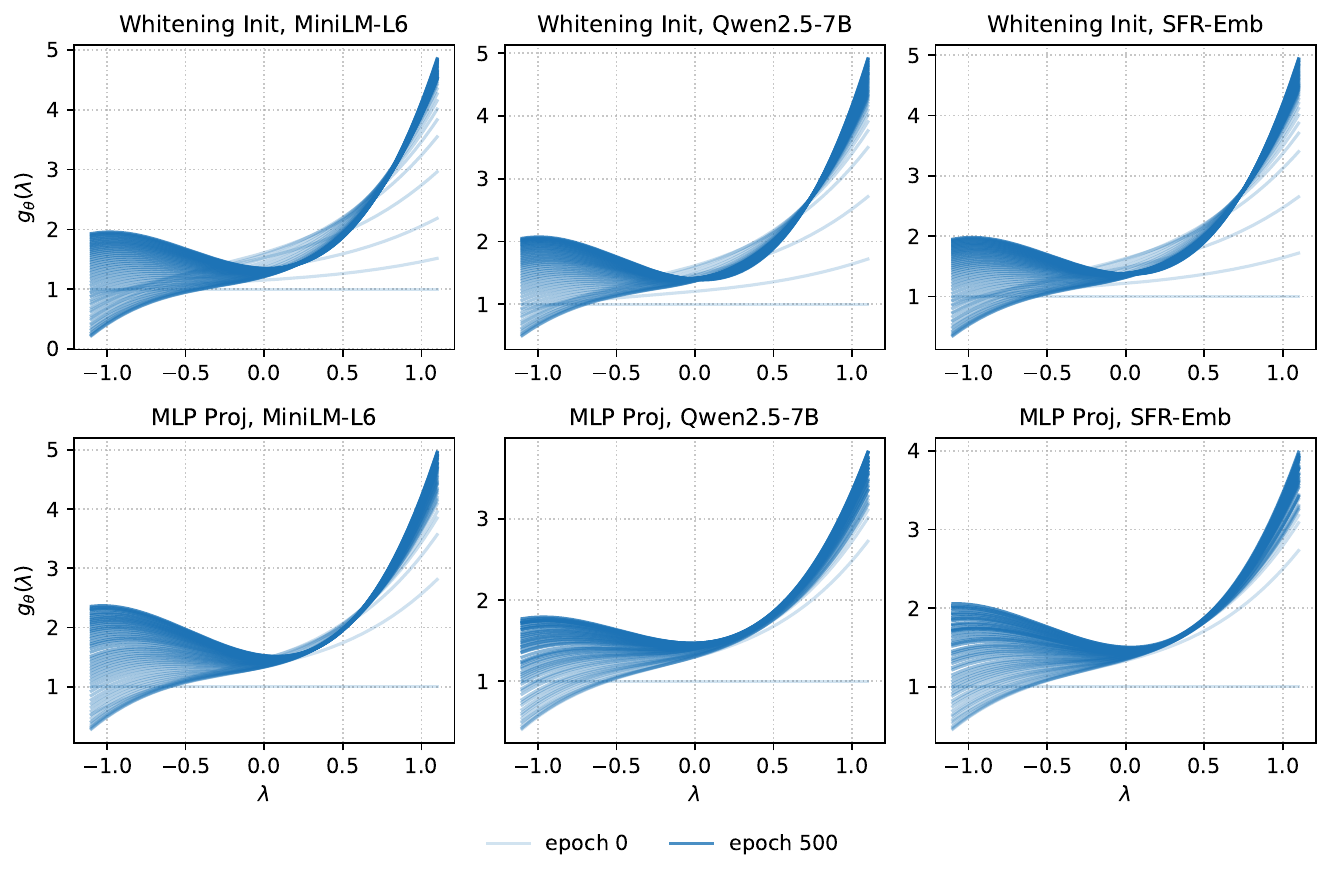}
  \caption{Filter Stability of ASPIRE in LLM-powered CF.}
  \label{figure-exp-llm-filter-stab-aspire-baby}
\end{figure}

For Naive-L (Figure~\ref{figure-exp-llm-filter-stab-naivel-baby}) equipped with the SFR-Emb encoder and MLP Proj adapter, the evolutionary pattern of low-frequency components exhibits slight discrepancies. 
Nevertheless, the prevalent low-frequency explosion issue remains prominent. 
Beyond this minor difference, both Naive-L and ASPIRE (Figure~\ref{figure-exp-llm-filter-stab-aspire-baby}) yield consistent overall behaviors with the conventional CF settings analyzed before.

\paragraph{Results on Electronics.}

\begin{table*}[t]
\centering

\caption{Performance comparison of LLM-assisted CF on Electronics. Best results are in \textbf{bold}.}

\begin{threeparttable}
\scalebox{0.825}{
\begin{tabular}{cccccccccccc}
\toprule
\multirow{2}{*}[-0.5ex]{Adapter} & \multirow{2}{*}[-0.5ex]{Filter} 
& \multirow{2}{*}[-0.5ex]{\makecell{Metric\\Sta.}}
& \multirow{2}{*}[-0.5ex]{\makecell{Filter\\Sta.}} 
& \multirow{2}{*}[-0.5ex]{\makecell{Filter\\Learn.}}
& \multicolumn{2}{c}{MiniLM-L6}
& \multicolumn{2}{c}{Qwen2.5-7B} 
& \multicolumn{2}{c}{SFR-Emb} \\
\cmidrule(lr){6-7} \cmidrule(lr){8-9} \cmidrule(lr){10-11}
& & & & & R@20 & N@20 & R@20 & N@20 & R@20 & N@20 \\
\midrule

\multirow{3}{*}{Whitening Init}
& Avg-P   & \xmark & \cmark & \xmark & 0.0649 & 0.0300 & 0.0635 & 0.0292 & 0.0655 & 0.0301 \\
& Naive-L & \xmark & \xmark     & \cmark & 0.0634 & 0.0288 & 0.0612 & 0.0277 & 0.0643 & 0.0293 \\
& ASPIRE  & \cmark & \cmark & \cmark & \textbf{0.0664} & \textbf{0.0301} & \textbf{0.0653} & \textbf{0.0297} & \textbf{0.0667} & \textbf{0.0305} \\

\midrule

\multirow{3}{*}{MLP Proj}
& Avg-P   & \cmark & \cmark & \xmark & 0.0540 & 0.0241 & 0.0588 & 0.0262 & 0.0619 & 0.0278 \\
& Naive-L & \xmark     & \xmark     & \cmark & 0.0497 & 0.0221 & 0.0520 & 0.0230 & 0.0561 & 0.0252 \\
& ASPIRE  & \cmark & \cmark & \cmark & \textbf{0.0582} & \textbf{0.0258} & \textbf{0.0616} & \textbf{0.0278} & \textbf{0.0621} & \textbf{0.0280} \\

\bottomrule
\end{tabular}
}

\begin{tablenotes}
\begin{minipage}{0.95\textwidth}
\footnotesize
\item[1] Sta.~=~Stability, Learn.~=~Learnability.
\item[2] The MiniLM-L6 column corresponds to all-MiniLM-L6-v2~\citep{wang2020minilm}, as adopted in Section~\ref{section-experiment}. Though strictly speaking this model is not an LLM, it is included for comparison. Qwen2.5-7B is officially released by \citet{qwen2.5}, and SFR-Emb denotes SFR-Embedding-Mistral~\citep{sfr-emb}.
\end{minipage}
\end{tablenotes}

\end{threeparttable}

\label{table-appendix-llm-cf}
\end{table*}

We also report results on Electronics in Table~\ref{table-appendix-llm-cf} for reference.

\subsection{Extended Sensitivity Analysis for Table~\ref{table-initialization-comparison}}
\label{appendix-init-sensitivity}

\begin{table}[]
  \centering
  \caption{
    Initialization Robustness on Ciao.
  }
  \label{table-appexdix-filter-init}
  
\scalebox{0.95}{
\begin{tabular}{lccccccc}
\toprule
\multirow{2}{*}{Initialization} & \multirow{2}{*}{$g_\theta^{init}(\lambda)$} 
& \multicolumn{2}{c}{Convergence Speed} 
& \multicolumn{4}{c}{Final Performance} \\
\cmidrule(lr){3-4} \cmidrule(lr){5-8}
& & E@80\% & E@95\% & R@10 & R@20 & N@10 & N@20 \\
\midrule
full-pass      & $1$ & 10  & 35 & 0.0489 & 0.0796 & 0.0309 & 0.0397 \\
zero-crossing  & $\lambda$ & 5   & 30  & 0.0488 & 0.0795 & 0.0309 & 0.0397 \\
low-pass       & $\frac{1}{4}\sum_{l=0}^{3}\lambda^{l}$ & 5 & 15 & 0.0490 & 0.0795 & 0.0309 & 0.0397 \\
mid-pass       & $1-\lambda^2$ & 15 & 30 & 0.0482 & 0.0790 & 0.0308 & 0.0396 \\
high-pass      & $1-2\lambda+\lambda^2$ & 15 & 45 & 0.0491 & 0.0797 & 0.0311 & 0.0398 \\
\midrule
\multicolumn{2}{c}{Baseline: Average-Pooling} & 5 & 60 & 0.0473 & 0.0770 & 0.0295 & 0.0381 \\
\bottomrule
\end{tabular}
}
\end{table}

\begin{figure}[]
  \centering
  \hspace*{0.0001\textwidth}
  \subfloat[(a) Baby]{\includegraphics[width=0.45\textwidth]{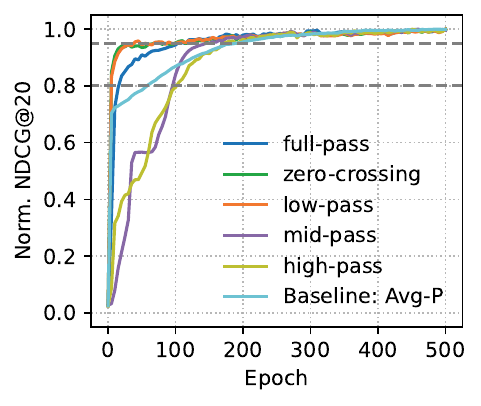}} \hfill
  \subfloat[(b) Ciao]{\includegraphics[width=0.45\textwidth]{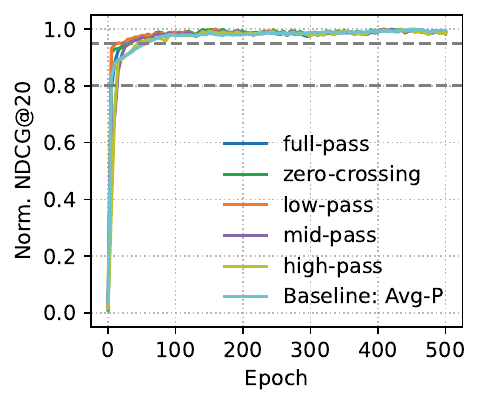}} \hfill
  \hspace*{0.0001\textwidth} 
\caption{
    Min-Max Normalized NDCG@20 Trajectory of ASPIRE under different initialization.
}
\label{fig-exp-init-convspeed}
\end{figure}

\begin{figure}
  \centering
  \includegraphics[width=0.75\textwidth]{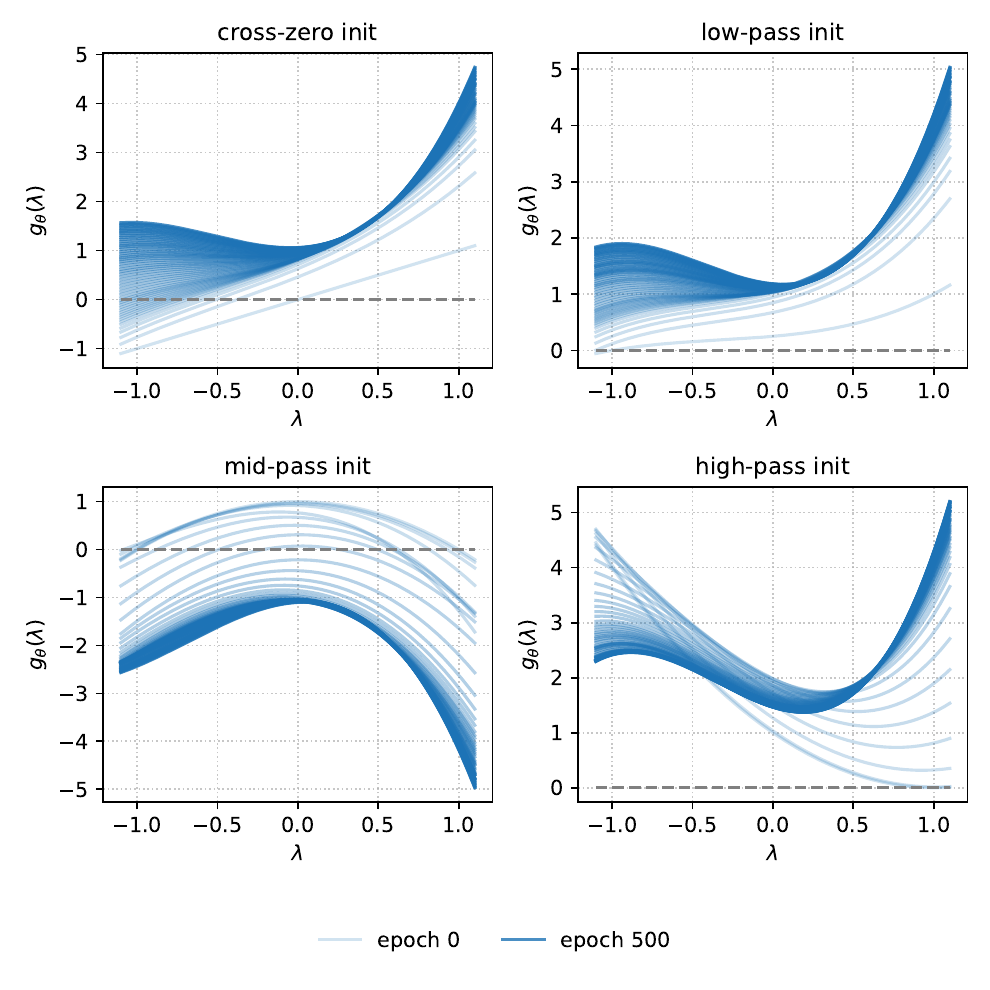}
  \caption{Filter evolution of ASPIRE under different initialization on Baby.}
  \label{figure-exp-init-filter-baby}
\end{figure}

\begin{figure}
  \centering
  \includegraphics[width=0.75\textwidth]{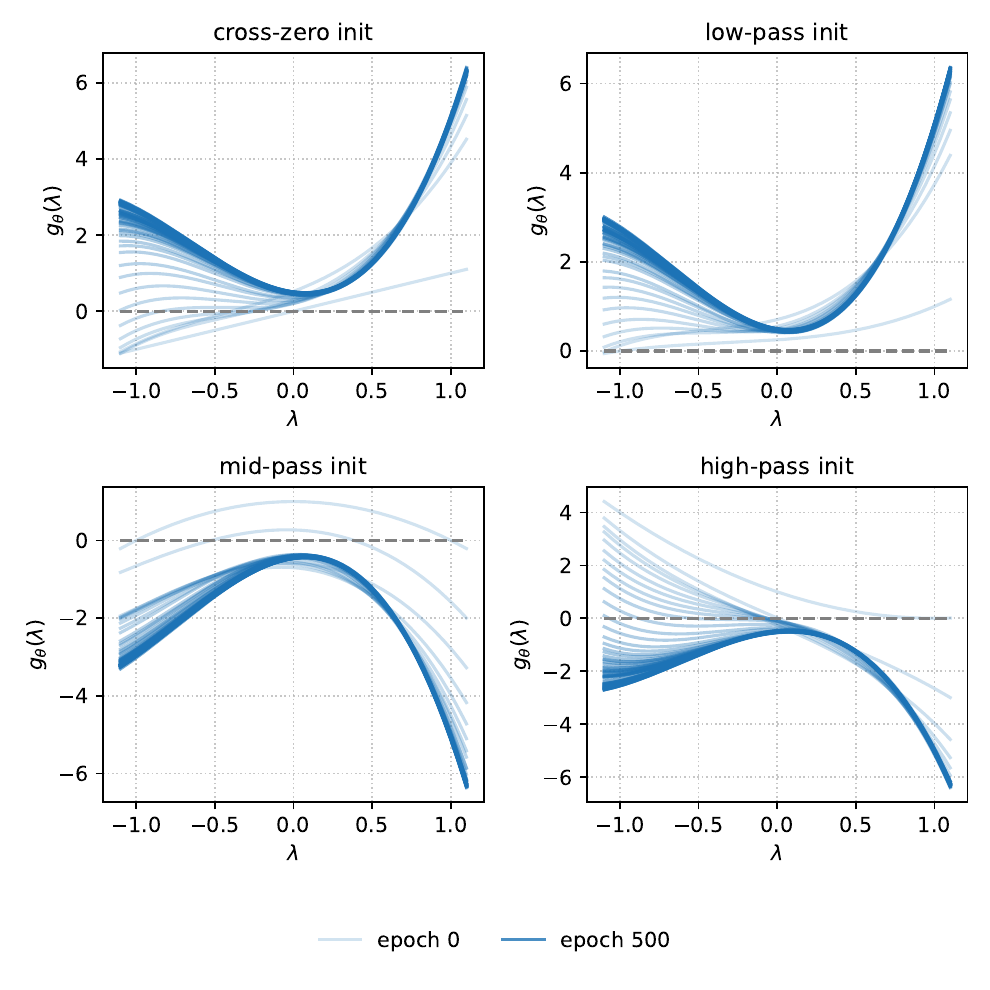}
  \caption{Filter evolution of ASPIRE under different initialization on Ciao.}
  \label{figure-exp-init-filter-ciao}
\end{figure}

We further conduct identical experiments on Ciao and present the corresponding results in Table~\ref{table-appexdix-filter-init}.
The convergence trends are highly consistent with the analysis elaborated in the main text.
In Figure~\ref{fig-exp-init-convspeed}, we normalize all NDCG@20 curves to the range [0,1] to intuitively compare the convergence speed of different initialization strategies.
The gray horizontal lines mark the 80\% and 95\% convergence thresholds reported in our quantitative tables.
From the visualized results, we conclude that prior-knowledge-aligned initializations (\eg, low-pass initialization) converge rapidly and possess prominent advantages over the Avg-P baseline.
In contrast, improper initializations that contradict inherent prior knowledge (\eg, high-pass initialization) degrade model performance in the early training phase, but such methods can quickly recover, and their overall convergence speeds are no worse than that of Avg-P.
As a neutral scheme, full-pass initialization delivers considerably faster convergence compared with Avg-P, though it still falls slightly behind the optimal prior-aware initializations.
In terms of the final performance, the five investigated initialization settings produce nearly indistinguishable results.
Additionally, the filters learned under different initializations share great similarity, as illustrated in Figure~\ref{figure-exp-init-filter-baby} and Figure~\ref{figure-exp-init-filter-ciao}.

\subsection{Effect of Pre-filter Normalization}
\label{appendix-effect-norm}

\begin{figure}[]
  \centering
  \hspace*{0.0001\textwidth}
  \subfloat[(a) Baby]{\includegraphics[width=0.45\textwidth]{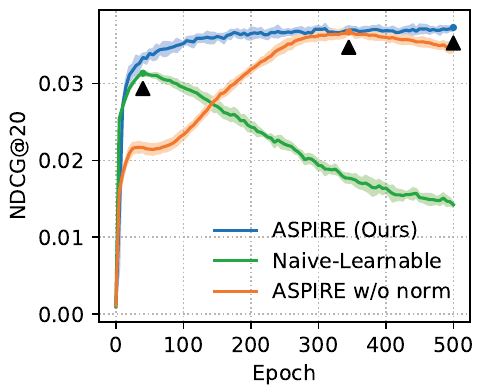}} \hfill
  \subfloat[(b) Ciao]{\includegraphics[width=0.45\textwidth]{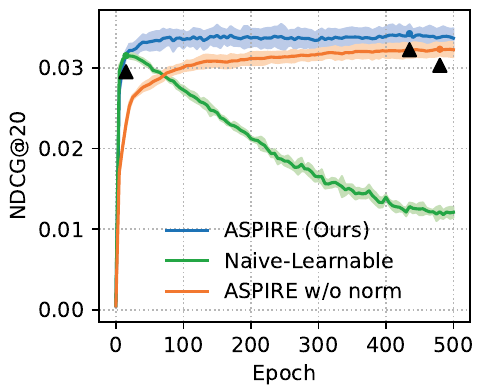}} \hfill
  \hspace*{0.0001\textwidth} 
\caption{
    NDCG@20 Trajectory comparison.
}
\label{fig-exp-metric-wo-prenorm}
\end{figure}

\begin{figure}
  \includegraphics[width=0.99\textwidth]{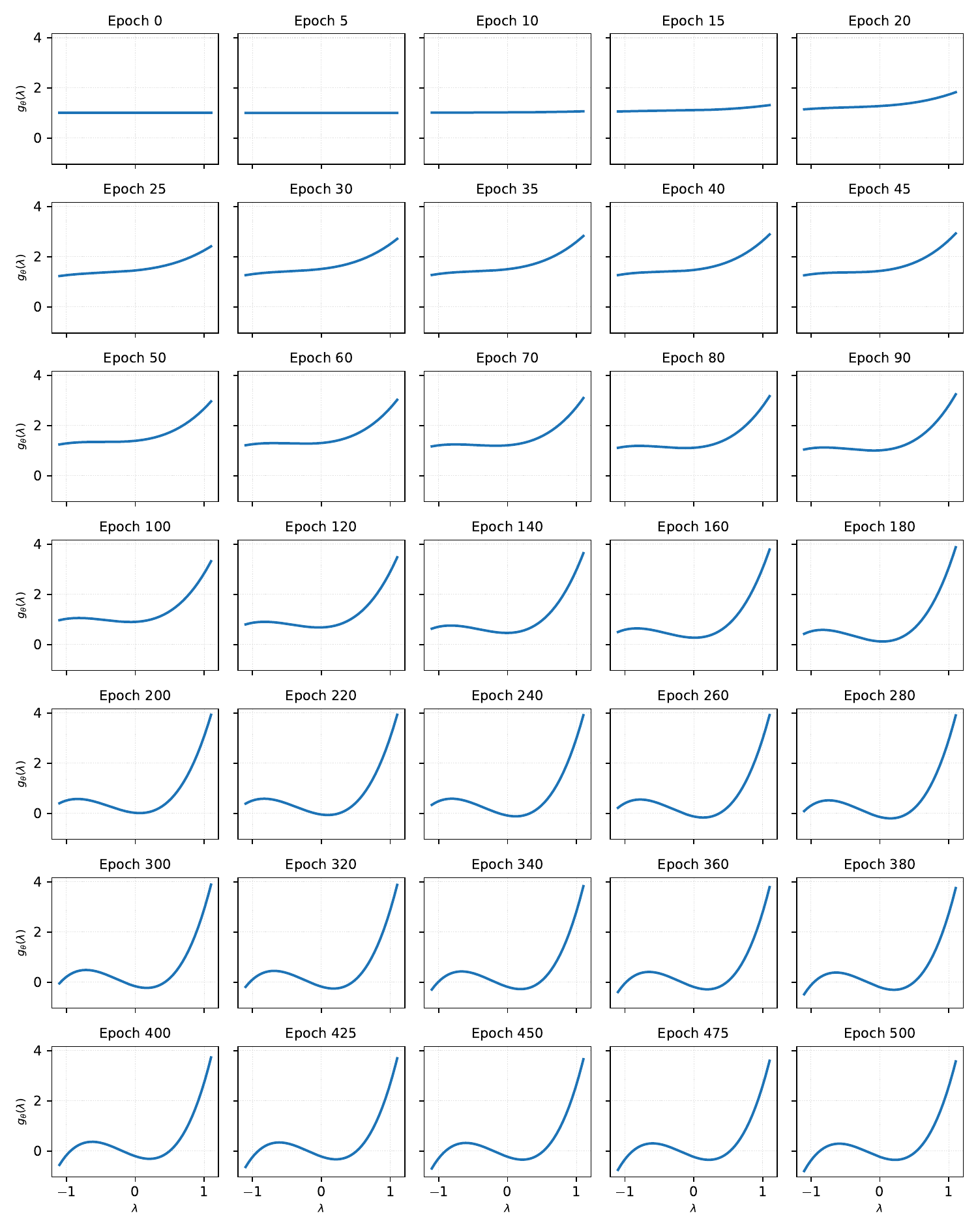}
  \caption{Filter evolution of ASPIRE without prefilter normalization on Baby.}
  \label{figure-exp-filter-evo-ablation-aspire-baby-full}
\end{figure}

\begin{figure}
  \includegraphics[width=0.99\textwidth]{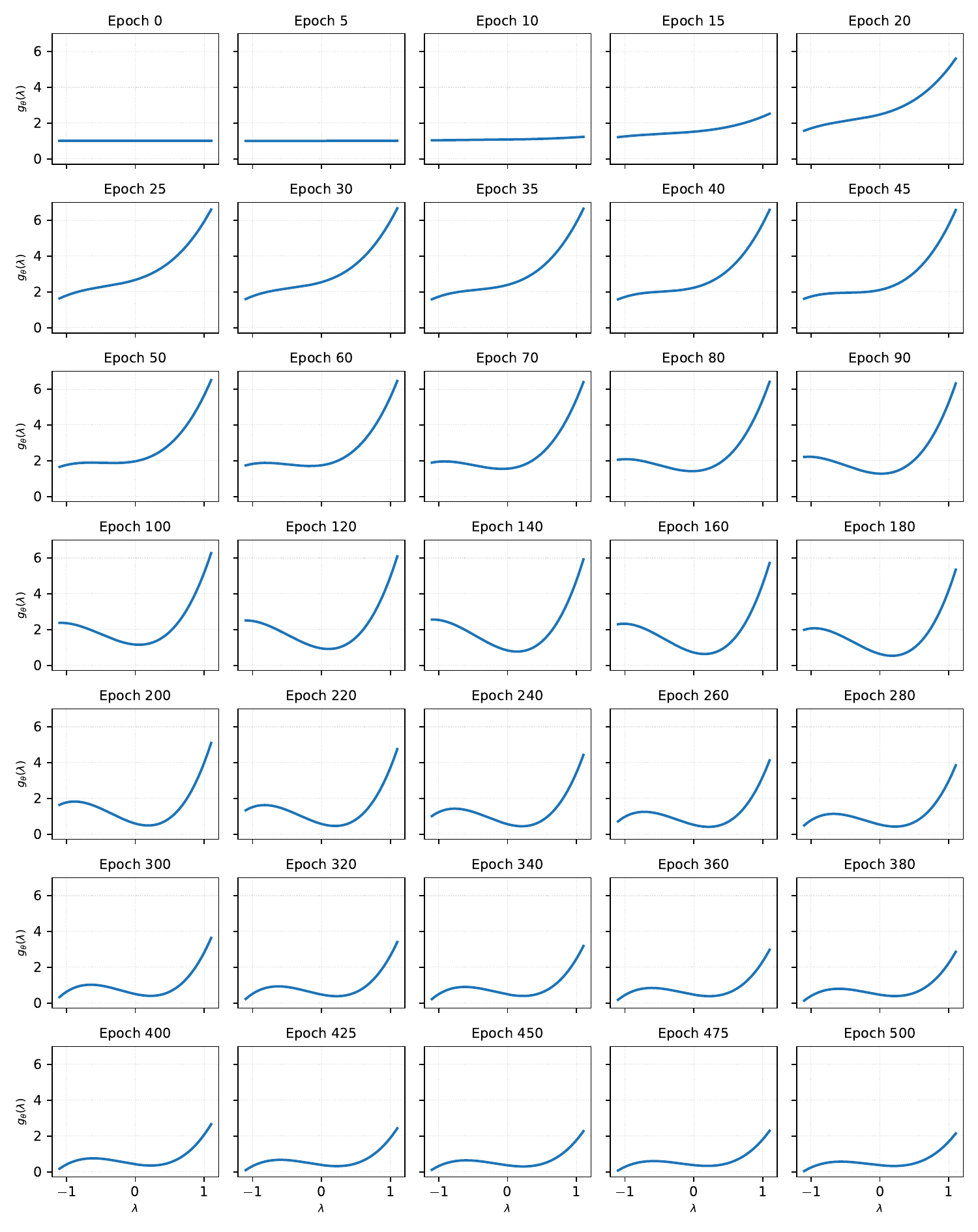}
  \caption{Filter evolution of ASPIRE without prefilter normalization on Ciao.}
  \label{figure-exp-filter-evo-ablation-aspire-ciao-full}
\end{figure}

\begin{figure}
  \includegraphics[width=0.99\textwidth]{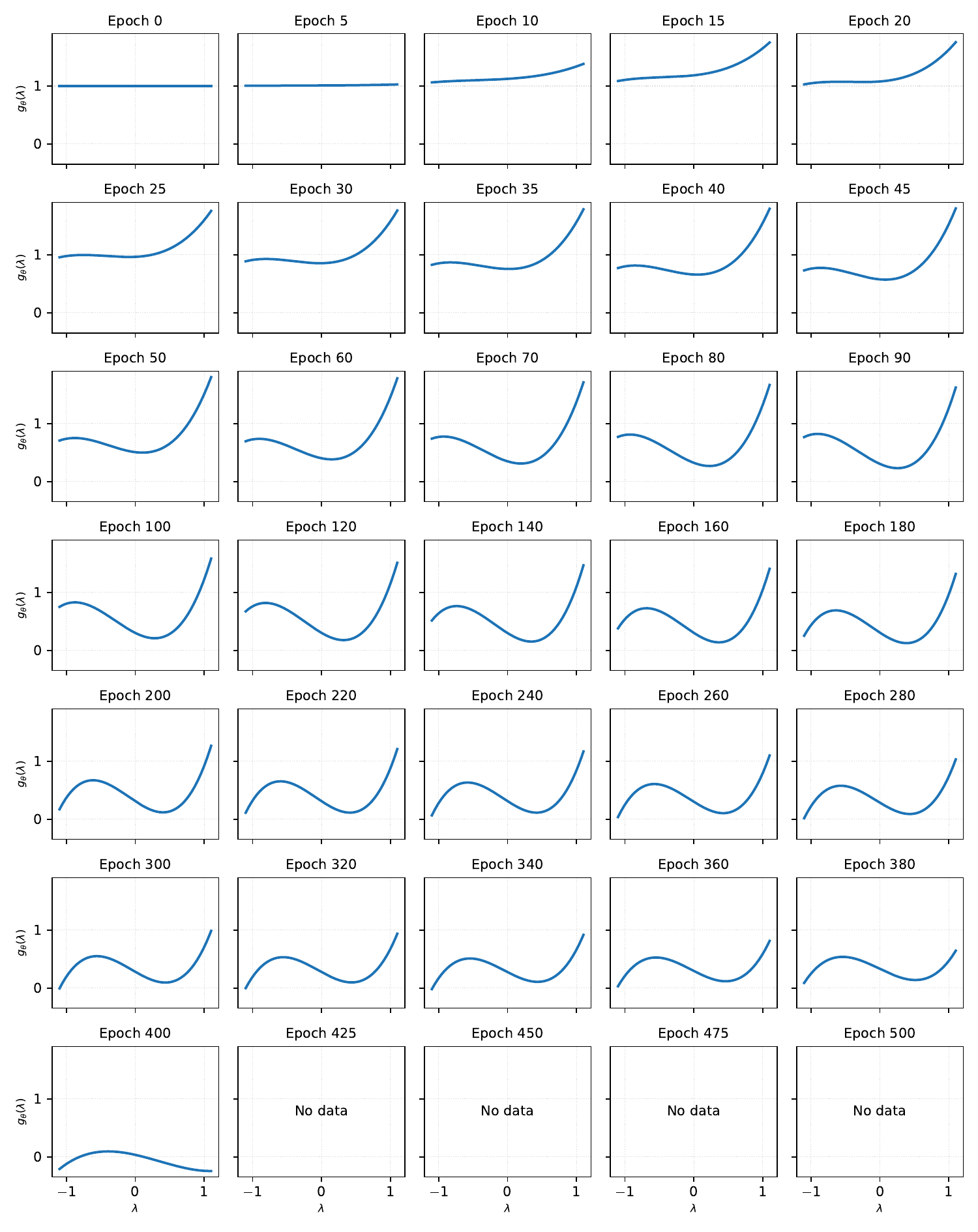}
  \caption{A bad case on Baby.}
  \label{figure-exp-filter-evo-ablation-aspire-baby-failure-full}
\end{figure}

As illustrated in Figure~\ref{fig-exp-metric-wo-prenorm}, the ASPIRE variant without pre-filter normalization yields inferior recommendation accuracy and exhibits considerably slower convergence. 
Besides, this model variant cannot guarantee stable training metrics, as validated by the filter evolution in Figure~\ref{figure-exp-filter-evo-ablation-aspire-baby-full} and Figure~\ref{figure-exp-filter-evo-ablation-aspire-ciao-full}. 
More critically, it is highly sensitive to hyperparameter configurations, where small deviations can lead to progressive filter degradation and eventually trigger a sudden loss divergence, causing training to collapse.
We select a representative bad case and visualize its filter evolution dynamics in Figure~\ref{figure-exp-filter-evo-ablation-aspire-baby-failure-full}.

\subsection{Time and Memory Efficiency}
\label{appendix-time-memory}

\begin{table*}[t]
\centering
\caption{Comparison of time and memory costs for different upper-level gradient approximations (memory in MB). N=Neumann, D=DARTS, Ex=Explicit Term, Im=Implicit Term. N-Ex is employed in ASPIRE. Results are obtained on an Intel Xeon E5-2680 v4 CPU and an NVIDIA RTX 3090 GPU.}
\label{table-appendix-cost-full}
\scalebox{.8}{
\setlength{\tabcolsep}{1.38mm}
\begin{tabular}{c | c | cccc | cccc | cccc}
\toprule
& \multirow{2}{*}{Method} 
& \multicolumn{4}{c|}{Baby} 
& \multicolumn{4}{c|}{Electronics} 
& \multicolumn{4}{c}{Ciao} \\
& & R@20 & N@20 & Time & Mem 
  & R@20 & N@20 & Time & Mem 
  & R@20 & N@20 & Time & Mem \\
\midrule

\multirow{6}{*}{\rotatebox{90}{Homogeneous}}
& Avg-P 
& 0.0851 & 0.0372 & 10m09s & 488 
& 0.0579 & 0.0272 & 120m16s & 1970 
& 0.0770 & 0.0381 & 9m53s & 640 \\

& Naive-L 
& 0.0690 & 0.0307 & 10m10s & 512 
& 0.0463 & 0.0213 & 120m11s & 1970 
& 0.0678 & 0.0352 & 9m52s & 640 \\

& N-Ex\&Im 
& 0.0870 & 0.0380 & 15m20s & 636 
& 0.0636 & 0.0306 & 281m30s & 3052 
& 0.0796 & 0.0397 & 14m23s & 644 \\

& \textbf{N-Ex} 
& 0.0871 & 0.0380 & 10m44s & 488 
& 0.0635 & 0.0306 & 139m38s & 2036 
& 0.0796 & 0.0397 & 10m37s & 640 \\

& D-Ex\&Im 
& 0.0873 & 0.0381 & 13m40s & 514 
& 0.0644 & 0.0310 & 194m27s & 2038 
& 0.0798 & 0.0398 & 13m25s & 640 \\

& D-Ex 
& 0.0877 & 0.0381 & 11m36s & 488 
& 0.0638 & 0.0306 & 159m21s & 2036 
& 0.0798 & 0.0397 & 11m34s & 640 \\

\midrule\midrule

\multirow{6}{*}{\rotatebox{90}{Heterogeneous}}
& Avg-P 
& 0.0990 & 0.0430 & 10m08s & 506 
& 0.0675 & 0.0313 & 159m47s & 2000 
& 0.0808 & 0.0410 & 9m30s & 658 \\

& Naive-L 
& 0.0833 & 0.0369 & 10m14s & 506 
& 0.0486 & 0.0228 & 159m36s & 2000 
& 0.0708 & 0.0360 & 9m37s & 658 \\

& N-Ex\&Im 
& 0.1029 & 0.0454 & 15m31s & 650 
& 0.0681 & 0.0314 & 365m17s & 3332 
& 0.0848 & 0.0429 & 14m40s & 664 \\

& \textbf{N-Ex} 
& 0.1027 & 0.0454 & 11m05s & 506 
& 0.0679 & 0.0313 & 181m50s & 2066 
& 0.0844 & 0.0426 & 10m26s & 658 \\

& D-Ex\&Im 
& 0.1031 & 0.0454 & 14m08s & 508 
& 0.0680 & 0.0314 & 251m09s & 2068 
& 0.0842 & 0.0427 & 12m47s & 658 \\

& D-Ex 
& 0.1035 & 0.0456 & 12m41s & 506 
& 0.0679 & 0.0314 & 209m15s & 2066 
& 0.0844 & 0.0426 & 11m10s & 658 \\

\bottomrule
\end{tabular}
}
\end{table*}

Two types of approximations are commonly employed to solve bilevel optimization problems.
The first is the Neumann approximation, which is adopted by~\citet{chen2023biao, yin2024dr4sr}. Specifically, the inverse Hessian term in ~\eqref{equation-neumann-approx} is approximated using the truncated Neumann series as
\begin{equation}
\left( \nabla_E^2 \mathcal{L}_{\mathrm{train}} \right)^{-1}
\approx
\sum_{m=0}^{M} \left( I - \nabla_E^2 \mathcal{L}_{\mathrm{train}} \right)^m.
\end{equation}
This approximation avoids the explicit computation of the Hessian inverse and enables efficient estimation via iterative Hessian-vector products.
The second is the DARTS approximation~\citep{liu2018darts}, used by~\citet{yu2024autognn, zhang2025bifair}:
\begin{equation}
\begin{gathered}
\nabla_\theta \mathcal{L}_{\mathrm{valid}}\left( \theta , E^*(\theta) \right) \approx 
\underbrace{
    \nabla_\theta \mathcal{L}_{\mathrm{valid}}\left( \theta, E^{\prime}\right)
}_{\text{Explicit Term}} 
\underbrace{
    - \frac{\xi}{2\epsilon}
    \left[
    \nabla_\theta \mathcal{L}_{\mathrm{train}}\left( \theta, E^+\right)
    -
    \nabla_\theta \mathcal{L}_{\mathrm{train}}\left( \theta, E^-\right)
    \right]
}_{\text{Implicit Term}}
\\
E^{\prime} = E - \xi \nabla_E \mathcal{L}_{\mathrm{train}}\left( \theta, E\right)
,\quad
E^{\pm} = E \pm \epsilon \nabla_E \mathcal{L}_{\mathrm{valid}}\left( \theta, E^{\prime}\right),
\end{gathered}
\end{equation}
where $\xi$ denotes the learning rate for the lower-level optimization, and $\epsilon$ is a small scalar controlling the finite-difference approximation. 
We compare these approximations in Table~\ref{table-appendix-cost-full}. 
ASPIRE and its three variants yield nearly identical recommendation performance.
However, the Neumann approximation without the implicit term is the most time- and memory-efficient, particularly on large-scale datasets (Electronics).
This comparison is conducted under a unified setting of 500 epochs, and when taking into account the accelerated convergence of ASPIRE, its actual convergence time is expected to be more advantageous than that of traditional methods.

\subsection{Ablation Study}
\label{appendix-ablation}

\begin{table*}[t]
\centering

\caption{Ablation study. Best results are in \textbf{bold}. PN, BO, and OS denote pre-filter normalization, bi-level objective, and oracle supervision, respectively. The effect of removing pre-filter normalization is dataset-dependent: it causes clear metric instability on Baby but remains relatively stable on Ciao. }

\scalebox{0.9}{
\begin{tabular}{ccccccccccc}
\toprule
\multirow{2}{*}[-0.5ex]{Variant}
& \multirow{2}{*}[-0.5ex]{\makecell{Metric\\Sta.}}
& \multirow{2}{*}[-0.5ex]{\makecell{Filter\\Sta.}}
& \multicolumn{4}{c}{Baby}
& \multicolumn{4}{c}{Ciao} \\
\cmidrule(lr){4-7} \cmidrule(lr){8-11}
& & 
& R@10 & R@20 & N@10 & N@20
& R@10 & R@20 & N@10 & N@20 \\
\midrule

Naive-L
& \xmark
& \xmark
& 0.0450
& 0.0690
& 0.0246
& 0.0307
& 0.0445
& 0.0678
& 0.0283
& 0.0352 \\

ASPIRE
& \cmark
& \cmark
& \textbf{0.0550}
& \textbf{0.0871}
& \textbf{0.0297}
& \textbf{0.0380}
& \textbf{0.0489}
& \textbf{0.0796}
& \textbf{0.0309}
& \textbf{0.0397} \\

w/o PN
& \cmark / \xmark
& \xmark
& 0.0526
& 0.0824
& 0.0286
& 0.0362
& 0.0470
& 0.0740
& 0.0298
& 0.0377 \\

w/o BO
& \xmark
& \xmark
& 0.0519
& 0.0815
& 0.0280
& 0.0356
& 0.0476
& 0.0731
& 0.0300
& 0.0375 \\

w/o OS
& \xmark
& \xmark
& 0.0363
& 0.0570
& 0.0200
& 0.0254
& 0.0279
& 0.0458
& 0.0189
& 0.0226 \\

\bottomrule
\end{tabular}
}

\label{table-ablation}
\end{table*}

Appendix~\ref{appendix-effect-norm} already studies the effect of pre-filter normalization, and we now demonstrate additional ablations that separately evaluate normalization, the bi-level objective, and oracle supervision in Table~\ref{table-ablation}.
From the experiment, we have the following observations:
\begin{itemize}
    \item Removing pre-filter normalization significantly reduces both performance and stability, consistent with Appendix~\ref{appendix-effect-norm}.
    \item Normalization alone can partially stabilize training, but low-frequency explosion and non-convergence still persist, and the resulting performance remains below most baselines, especially Avg-P.
    \item Replacing the validation-set oracle supervision with training-set supervision, while retaining the other designs, makes the instability more severe and substantially degrades recommendation performance.
\end{itemize}

\subsection{Hyperparameter Sensitivity Analysis}
\label{appendix-sensitivity}

\paragraph{Filter Order / Convolution Layers $L$.}

\begin{figure}
  \centering
  \includegraphics[width=0.75\textwidth]{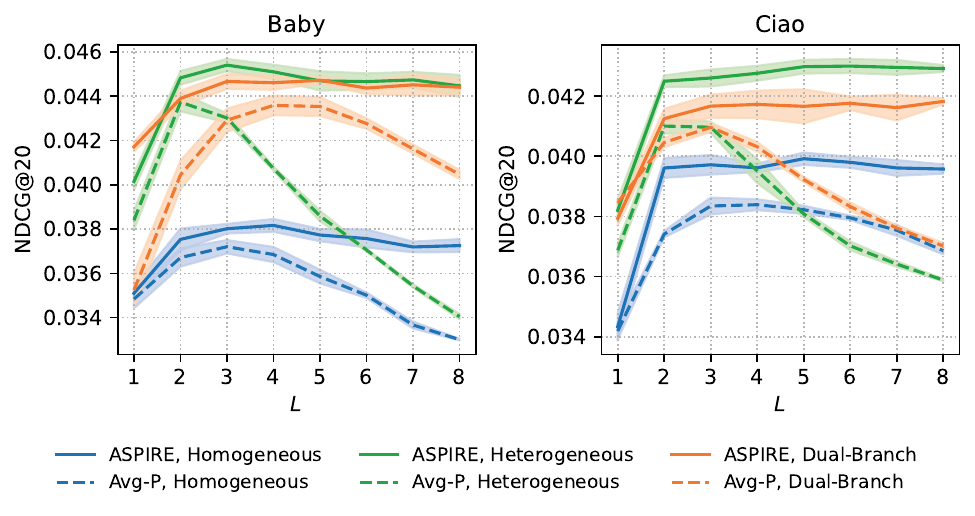}
  \caption{Performance comparison between ASPIRE and Avg-P with different layers.}
  \label{figure-exp-layers-perf}
\end{figure}

\begin{figure}
  \centering
  \includegraphics[width=0.75\textwidth]{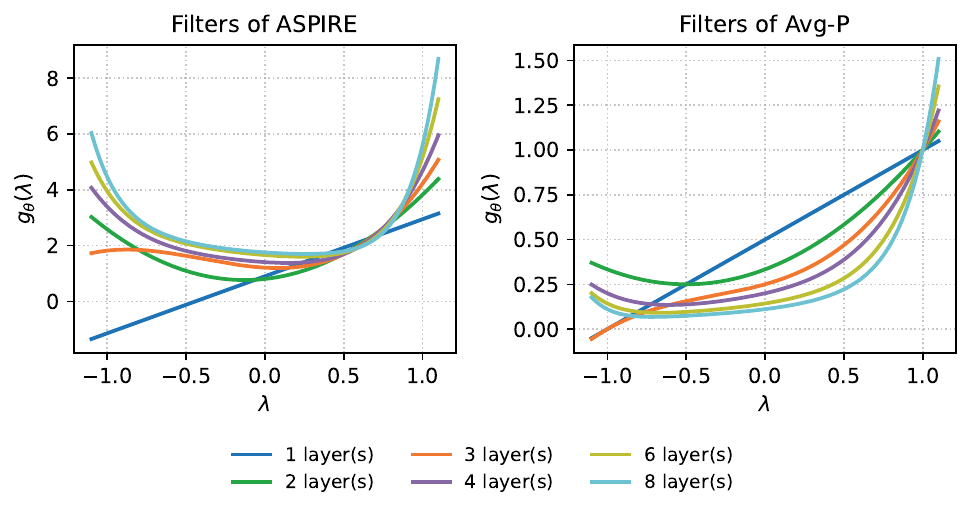}
  \caption{Filter comparison between ASPIRE and Avg-P with different layers.}
  \label{figure-exp-layers-filter}
\end{figure}

In this experiment, we vary the number of propagation layers and compare ASPIRE with Avg-P, as shown in Figure~\ref{figure-exp-layers-perf}.
For Avg-P, the best performance typically occurs around 2-4 layers and degrades noticeably as the depth increases, a phenomenon commonly attributed to over-smoothing~\citep{liu2021impgcn, wu2024afdgcf, li2018oversmoothing}.
In contrast, ASPIRE exhibits significantly more stable performance once the number of layers exceeds three.
To better understand this difference, we visualize the learned spectral filters in Figure~\ref{figure-exp-layers-filter}. 
As the number of layers increases, the filters learned by ASPIRE gradually converge to a consistent shape, indicating that higher-order representations are not strictly necessary but are also not detrimental.
By comparison, Avg-P increasingly biases the spectral response toward low-frequency components, progressively suppressing higher-frequency signals as the depth grows.
This demonstrates that ASPIRE not only effectively mitigates over-smoothing, but also holds strong promise as a practical tool for graph spectral analysis.

\paragraph{Temperature $\tau$.}

\begin{figure}
  \centering
  \includegraphics[width=0.75\textwidth]{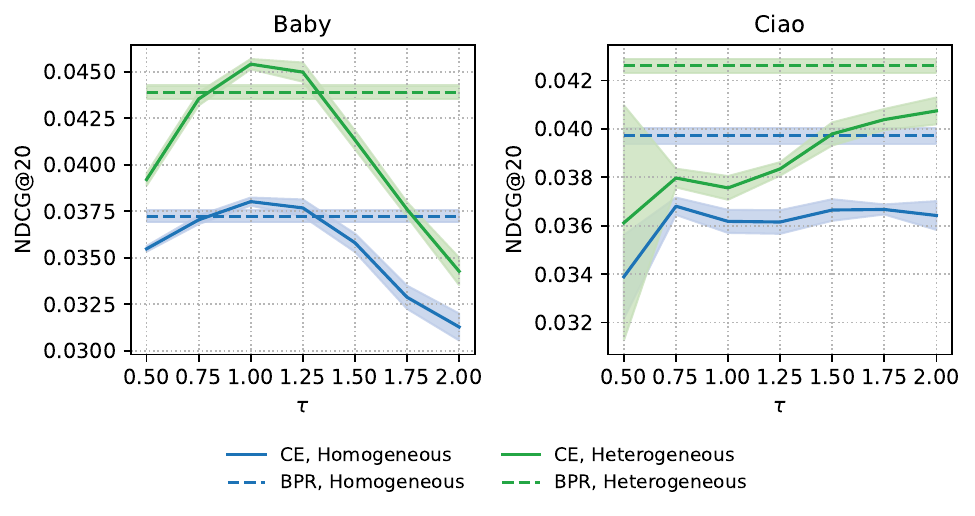}
  \caption{Performance of ASPIRE with different temperature settings in $\mathcal{L}_{\mathrm{valid}}$.}
  \label{figure-exp-temperatures}
\end{figure}

According to Appendix~\ref{appendix-implementation}, we evaluate both BPR and CE as $\mathcal{L}_{\mathrm{valid}}$. 
Specifically, CE performs better on the Baby dataset, whereas BPR achieves better results on Ciao (Figure~\ref{figure-exp-temperatures}).

\paragraph{Upper-level Optimization Configurations.}

\begin{figure}
  \centering
  \includegraphics[width=0.75\textwidth]{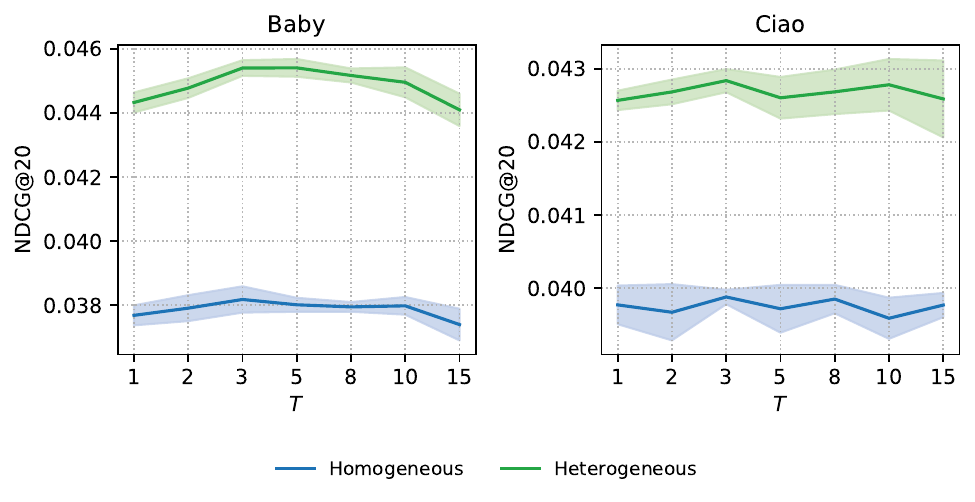}
  \caption{Performance of ASPIRE with different upper-level update intervals.}
  \label{figure-exp-sensi-interval}
\end{figure}

\begin{figure}
  \centering
  \includegraphics[width=0.75\textwidth]{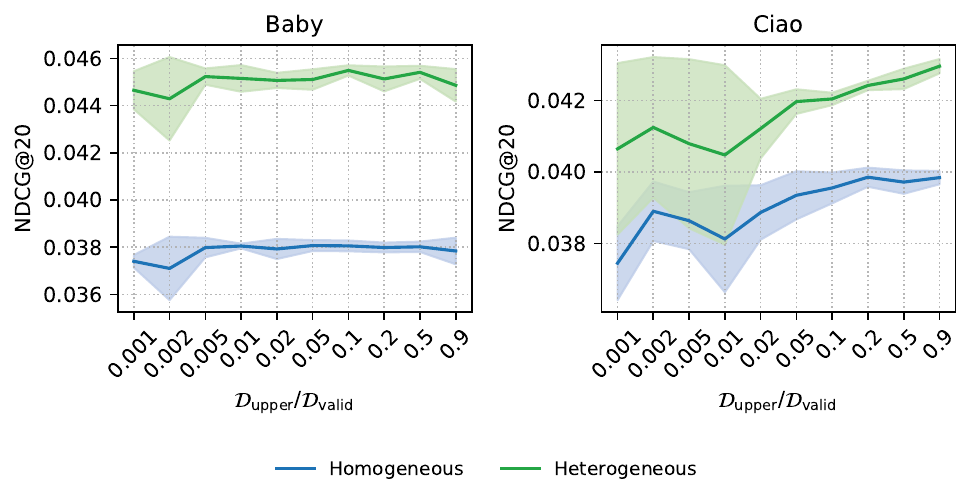}
  \caption{Performance of ASPIRE with different dataset sizes for upper-level updates.}
  \label{figure-exp-sensi-valratio}
\end{figure}

The upper-level update interval $T$ and the dataset size are not critical to the performance of ASPIRE. As illustrated in Figure~\ref{figure-exp-sensi-interval}, a moderate choice of $T$ yields slightly better results, while both overly frequent and overly infrequent updates lead to only marginal differences.
Figure~\ref{figure-exp-sensi-valratio} shows that on Baby, as little as 0.5\% of the validation set is sufficient for effective upper-level optimization.
On Ciao, a more noticeable performance drop, along with increased variance across random seeds, is observed when the ratio falls below 5\%.

\section{Broader Impact and Limitations}
\label{appendix-limitations}

Given the advantages of ASPIRE demonstrated in Section~\ref{section-experiment} and Section~\ref{appendix-sensitivity}, 
we believe it has strong potential as a practical tool for graph spectral analysis.
Existing spectral analyses~\citep{guo2023jgcf, he2025ssc} for a given graph are typically restricted to specific graph classes and mainly provide qualitative characterizations of frequency distributions. 
They rarely offer an accurate estimate of the effective spectral filter that a learning objective induces. 
In contrast, ASPIRE directly learns a quantitative filter function from data, providing an effective and precise approximation of the spectral response preferred by the interaction graph.

Another long-standing problem in collaborative filtering is the parameterization of propagation layers, \eg, introducing linear transformations between layers to enhance expressivity~\citep{wang2019ngcf}.
Such parameterizations also suffer from abrupt performance degradation, consistent with the phenomenon illustrated in Figure~\ref{fig-intro-group}(a).
~\citet{xu2023stablegcn} attempted to address this issue, but their approach mainly increases the complexity of pre-filter encoding while still relying on a fixed graph filter as in LightGCN.
The stability of ASPIRE, as discussed in Section~\ref{section-stable-converge}, further highlights its potential to support the development of more expressive and stable graph neural networks for collaborative filtering.

However, this work still has several limitations.
From a spectral perspective on collaborative filtering, our experiments do not cover more complex GNN architectures, such as those incorporating learnable linear transformations.
Also, our exploration of LLM-powered collaborative filtering is limited to using LLMs as encoder.
It remains to be investigated whether the low frequency explosion phenomenon manifests across a wider range of tasks, such as sequential recommendation, and whether ASPIRE remains effective in those settings.


\section{LLM Usage}

Large language models were used solely for grammar checking and language polishing of the manuscript.

\end{document}

%% file: iclr2027/math_commands.tex
\usepackage{amsmath,amsfonts,bm}

\def\eqref#1{equation~\ref{#1}}

\def\1{\bm{1}}

\DeclareMathAlphabet{\mathsfit}{\encodingdefault}{\sfdefault}{m}{sl}
\SetMathAlphabet{\mathsfit}{bold}{\encodingdefault}{\sfdefault}{bx}{n}

\DeclareMathOperator*{\argmin}{arg\,min}